\documentclass[12pt,english]{article}
\usepackage[T1]{fontenc}
\usepackage[latin1]{inputenc}
\usepackage{geometry}
\usepackage{babel}
\usepackage{graphics}
\def\beq{\begin{equation}}
\def\eeq{\end{equation}}

\def\R{{\bf R}}
\def\beqn{\begin{eqnarray}}
\def\eeqn{\end{eqnarray}}
\def\ba{\begin{eqnarray}}
\def\ea{\end{eqnarray}}

\def\one {\bf  1}

\newcommand{\beqa}{\begin{eqnarray}}
\newcommand{\eeqa}{\end{eqnarray}}

\makeatletter

\providecommand{\LyX}{L\kern-.1667em\lower.25em\hbox{Y}\kern-.125emX\@}

\makeatother
\begin{document}
\begin{center}
\vspace{3.cm}
{\bf \Large Direct Solution of Renormalization Group Equations of QCD in $x-$space:\\}
{\bf \large  NLO Implementations at Leading Twist \\}
\vspace{2cm}
 {\large Alessandro Cafarella and Claudio Corian\`{o}\\}
\vspace{1.5cm}
{\it Dipartimento di Fisica, Università di Lecce and INFN sezione
di Lecce\\
Via per Arnesano, 73100 Lecce, Italy\footnote{Permanent Address}\\
and\\
Department of Physics and Institute for Plasma Physics, 
Univ. of Crete \\
and FO.R.T.H., 71003 Heraklion, Greece\\}
{\it e-mail: alessandro.cafarella@le.infn.it, claudio.coriano@le.infn.it}
\end{center}
\vspace{2.cm}
\begin{abstract}
We illustrate the implementation of a method based on the use of recursion
relations in (Bjorken) $x-$space 
for the solution of the evolution equations of QCD for all the leading twist distributions.  
The algorithm has the advantage of being very fast. The implementation 
that we release is written in C and is performed to next-to-leading 
order in $\alpha_s$.

\end{abstract}
\newpage
\section*{Program summary}

\begin{itemize}
\item \emph{Title of program:} \texttt{evolution.c}
\item \emph{Computer:} Athlon 1800 plus
\item \emph{Operating system under which the program has been tested:} Linux
\item \emph{Programming language used:} C
\item \emph{Peripherals used:} Laser printer
\item \emph{No. of bytes in distributed program:} 20771
\item \emph{Keywords:} Polarized and unpolarized 
parton distribution, numerical solutions of renormalization group equation 
in Quantum Chromodynamics.
\item \emph{Nature of physical problem:} The program provided here solves 
the DGLAP evolution equations to next-to-leading order \( \alpha _{s} \), for unpolarized, longitudinally polarized
and transversely polarized parton distributions.
\item \emph{Method of solution:} We use a recursive method based on an expansion
of the solution in powers of \( \log \left( \alpha _{s}(Q)/\alpha _{s}(Q_{0})\right)  \)
\item \emph{Typical running time:} About 1 minute and 30 seconds for the
unpolarized and longitudinally polarized cases and 1 minute for the
transversely polarized case.
\end{itemize}

\section*{LONG WRITE UP}

\section{Introduction}
Our understanding of the QCD dynamics has improved steadily along the years. 
In fact we can claim that precision studies of the QCD background in a variety of 
energy ranges, from intermediate to high energy - 
whenever perturbation theory and factorization theorems hold - 
are now possible and the level of accuracy reached in this area is due both 
to theoretical improvements and to the flexible numerical implementations 
of many of the algorithms developed at theory level.  

Beside their importance in the determination of the QCD background in the search of new physics, 
these studies convey an understanding of the structure of the nucleon from first principles, 
an effort which is engaging both theoretically and experimentally.  

It should be clear, however, that perturbative methods 
are still bound to approximations, from the order of the perturbative 
expansion to the phenomenological description of the parton distribution functions, being them related to a parton model view of the QCD dynamics.

Within the framework of the parton model, evolution equations of DGLAP-type 
- and the corresponding initial conditions on the parton distributions -
are among the most important 
blocks which characterize the description of
the quark-gluon interaction and, as such, deserve continuous attention.
Other parts of this description require the computation of hard scatterings 
with high accuracy and an understanding of the fragmentation region as well.
The huge success of the parton model justifies all the effort. 

In this consolidated program, we believe that any attempt to study the renormalization
group evolution describing the perturbative change of the 
distributions with energy, from a different - in this case, numerical - standpoint is of interest.

In this paper we illustrate an algorithm based on the
use of recursion relations for the solution of evolution equations
of DGLAP type. We are going to discuss some of the salient features 
of the method and illustrate 
its detailed implementation as a computer code 
up to next-to leading order (NLO) in $\alpha_s$, the strong coupling constant. 

In this context, it is worth to recall that the most common method 
implemented so far in the solution of the DGLAP equations is the one based
on the Mellin moments, with all its good points and limitations. 

The reason for such limitations are quite obvious: while it is rather straightforward to solve
the equations in the space of moments, 
their correct inversion is harder to perform, since this 
requires the computation of an integral 
in the complex plane and the search for an optimized path.

In this respect, several alternative implementations of the NLO 
evolution are available from the previous literature, either based on
the use of ``brute force'' algorithms \cite{bruteforce} or 
on the Laguerre expansion \cite{Petronzio,CorianoSavkli}, all with their positive features and their limitations. 

Here we document an implementation to NLO of a method based on an
ansatz \cite{Rossi} which allows to rewrite the evolution
equations as a set of recursion relations for some scale invariant
functions, \( A_{n}(x) \) and \( B_{n}(x) \), which appear in the
expansion. The advantage, compared to others, 
of using these recursion relations is that just few iterates 
of these are necessary in order to obtain a stable solution. 
The number of iterates is determined at run time.
We also mention that 
our implementation can be extended to more complex cases, including 
the cases of nonforward parton distributions and of supersymmetry, 
which will be discussed elsewhere . Here 
we have implemented the recursion method in all the important cases of 
initial state scaling violations connected to the evolutions of both 
polarized and unpolarized parton densities, including the less known transverse spin distributions.

 Part of this document is supposed to illustrate the algorithm, having 
worked out in some detail the structure of the recursion relations rather explicitly, 
especially in the case of non-singlet evolutions, such as for transverse 
spin distributions. As we have mentioned, 
the same method has also been applied successfully 
to the case of supersymmetric QCD and will be illustrated in a companion paper. 

One of the advantages of the method is its analytical base,  
since the recursion relations 
can be written down in explicit form and at the same time 
one can perform a simple analytical matching between various regions in the 
evolutions, which is a good feature of x-spaced methods. 
While this last aspect is not 
relevant for ordinary QCD, it is relevant in other theories, such as 
for supersymmetric extensions of the parton model, once one assumes 
that, as the evolution scale $Q$ raises, new anomalous dimensions are needed 
in order to describe the mixing between ordinary and supersymmetric partons.

\section{Definitions and Conventions}

In this section we briefly outline our definitions and conventions.

We will be using the running of the coupling constant up to two-loop level\begin{equation}
\label{eq:alpha_s}
\alpha _{s}(Q^{2})=\frac{4\pi }{\beta _{0}}\frac{1}{\log (Q^{2}/\Lambda _{\overline{MS}}^{2})}\left[ 1-\frac{\beta _{1}}{\beta ^{2}_{0}}\frac{\log \log (Q^{2}/\Lambda _{\overline{MS}}^{2})}{\log (Q^{2}/\Lambda _{\overline{MS}}^{2})}+O\left( \frac{1}{\log ^{2}(Q^{2}/\Lambda _{\overline{MS}}^{2})}\right) \right] ,
\end{equation}
where\begin{equation}
\beta _{0}=\frac{11}{3}N_{C}-\frac{4}{3}T_{f},\qquad \beta _{1}=\frac{34}{3}N^{2}_{C}-\frac{10}{3}N_{C}n_{f}-2C_{F}n_{f},
\end{equation}
and\begin{equation}
N_{C}=3,\qquad C_{F}=\frac{N_{C}^{2}-1}{2N_{C}}=\frac{4}{3},\qquad T_{f}=T_{R}n_{f}=\frac{1}{2}n_{f},
\end{equation}
where \( N_{C} \) is the number of colors, \( n_{f} \) is the number
of active flavors, which is fixed by the number of quarks with \( m_{q}\leq Q \).
We have taken for the quark masses 
$ m_{c}=1.5\, \textrm{GeV}, m_{b}=4.5\, \textrm{GeV}$ 
and $ m_{t}=175\, \textrm{GeV} $, these are necessary in order 
to identify the thresholds at which the number of flavours $n_f$ 
is raised as we increase the final evolution scale.

 In our conventions 
$\Lambda_{QCD}$ is denoted by $\Lambda _{\overline{MS}}^{(n_{f})}$
and is given by

\begin{equation}
\Lambda _{\overline{MS}}^{(3,4,5,6)}=0.248,\, 0.200,\, 0.131,\, 0.050\, \textrm{GeV}.
\end{equation}

We also define the distribution of a given helicity $(\pm)$,
\( f^{\pm }(x,Q^{2}) \), which is the probability 
of finding a parton of type \( f \) at a scale \( Q \), 
where \( f=q_{i},\overline{q_{i}},g \),
in a longitudinally polarized proton with the spin aligned (+) 
or anti-aligned (-) respect 
to the proton spin and carrying a fraction \( x \) of the proton's
momentum. 

As usual, we introduce the longitudinally polarized parton
distribution of the proton\begin{equation}
\label{eq:long_distr}
\Delta f(x,Q^{2})\equiv f^{+}(x,Q^{2})-f^{-}(x,Q^{2}).
\end{equation}
We also introduce another type of parton density, termed 
{\em transverse spin 
distribution}, which is defined as 
the probability of finding a parton of type \( f \) in a transversely
polarized proton with its spin parallel ($\uparrow$) minus the probability 
of finding it antiparallel ($\downarrow$) to the proton
spin
\begin{equation}
\Delta _{T}f(x,Q^{2})\equiv f^{\uparrow }(x,Q^{2})-f^{\downarrow }(x,Q^{2}).
\end{equation}

Similarly, the unpolarized (spin averaged) parton distribution of the proton
is given by
\begin{equation}
\label{eq:unp_distr}
f(x,Q^{2})\equiv f^{+}(x,Q^{2})+f^{-}(x,Q^{2})=f^{\uparrow }(x,Q^{2})+f^{\downarrow }(x,Q^{2}).
\end{equation}
We also recall, if not obvious, that 
taking linear combinations of Equations (\ref{eq:unp_distr}) and
(\ref{eq:long_distr}), one recovers 
the parton distributions of a given helicity 
\begin{equation}
f^{\pm }(x,Q^{2})=\frac{f(x,Q^{2})\pm \Delta f(x,Q^{2})}{2}.
\end{equation}
In regard to the kernels, the notations \( P \), \( \Delta P \), \( \Delta _{T}P \),
\( P^{+\pm} \),  will be used to denote the Altarelli-Parisi
kernels in the unpolarized, longitudinally polarized, transversely
polarized, and the positive (negative) helicity cases respectively.

Let us now turn to the evolution equations, starting from the unpolarized
case. Defining\begin{equation}
\label{eq:definizioni}
q_{i}^{(\pm )}=q_{i}\pm \overline{q_{i}},\qquad q^{(+)}=\sum _{i=1}^{n_{f}}q_{i}^{(+)},\qquad \chi _{i}=q_{i}^{(+)}-\frac{1}{n_{f}}q^{(+)},
\end{equation}
and introducing the convolution product\begin{equation}
f(x)\otimes g(x)=\int _{x}^{1}\frac{\textrm{d}y}{y}f\left( \frac{x}{y}\right) g(y),
\end{equation}
the evolution equations are  
\begin{equation}
\frac{\textrm{d}}{\textrm{d}\log Q^{2}}q_{i}^{(-)}(x,Q^{2})=P_{NS^{-}}(x,\alpha _{s}(Q^{2}))\otimes q_{i}^{(-)}(x,Q^{2}),
\end{equation}
\begin{equation}
\frac{\textrm{d}}{\textrm{d}\log Q^{2}}\chi _{i}(x,Q^{2})=P_{NS^{+}}(x,\alpha _{s}(Q^{2}))\otimes \chi _{i}(x,Q^{2}),
\end{equation}
for the non-singlet sector and
\begin{equation}
\label{eq:singlet}
\frac{\textrm{d}}{\textrm{d}\log Q^{2}}\left( \begin{array}{c}
q^{(+)}(x,Q^{2})\\
g(x,Q^{2})
\end{array}\right) =\left( \begin{array}{cc}
P_{qq}(x,\alpha _{s}(Q^{2})) & P_{qg}(x,\alpha _{s}(Q^{2}))\\
P_{gq}(x,\alpha _{s}(Q^{2})) & P_{gg}(x,\alpha _{s}(Q^{2}))
\end{array}\right) \otimes \left( \begin{array}{c}
q^{(+)}(x,Q^{2})\\
g(x,Q^{2})
\end{array}\right) 
\end{equation}
for the singlet sector.
Equations analogous to (\ref{eq:definizioni}-\ref{eq:singlet}),
with just a change of notation, are valid in the longitudinally polarized
case and, due to the linearity of the evolution equations, also for
the distributions in the helicity basis. In the transverse case instead,
there is no coupling between gluons and quarks, so the singlet sector
(\ref{eq:singlet}) is missing. In this case we will solve just the
nonsinglet equations \begin{equation}
\frac{\textrm{d}}{\textrm{d}\log Q^{2}}\Delta _{T}q_{i}^{(-)}(x,Q^{2})=\Delta _{T}P_{NS^{-}}(x,\alpha _{s}(Q^{2}))\otimes \Delta _{T}q_{i}^{(-)}(x,Q^{2}),
\end{equation}
\begin{equation}
\frac{\textrm{d}}{\textrm{d}\log Q^{2}}\Delta _{T}q_{i}^{(+)}(x,Q^{2})=\Delta _{T}P_{NS^{+}}(x,\alpha _{s}(Q^{2}))\otimes \Delta _{T}q_{i}^{(+)}(x,Q^{2}).
\end{equation}

We also recall that the perturbative expansion, up to next-to-leading order, of the kernels
is\begin{equation}
P(x,\alpha _{s})=\left( \frac{\alpha _{s}}{2\pi }\right) P^{(0)}(x)+\left( \frac{\alpha _{s}}{2\pi }\right) ^{2}P^{(1)}(x)+\ldots .
\end{equation}

Kernels of fixed helicity can also be introduced
with $P_{++}(z)=(P(z)+\Delta P(z))/2$ and $ P_{+-}(z)=(P(z)-\Delta P(z))/2$ 
denoting splitting functions of fixed helicity, which will be used below.

 The equations, in the helicity basis, are 

\begin{eqnarray}
{dq_+(x) \over{dt}}=
{\alpha_s \over {2 \pi}} (P_{++}^{qq} ({x \over y}) \otimes q_+(y)+
P_{+-}^{qq} ({x \over y}) \otimes q_-(y)  \nonumber \\
+P_{++}^{qg} ({x \over y}) \otimes g_+(y)+
P_{+-}^{qg} ({x \over y}) \otimes g_-(y)),
\nonumber \\
{dq_-(x) \over{dt}}=
{\alpha_s \over {2 \pi}} (P_{+-} ({x \over y}) \otimes q_+(y)+
P_{++} ({x \over y}) \otimes q_-(y) \nonumber \\
+P_{+-}^{qg} ({x \over y}) \otimes g_+(y)+
P_{++}^{qg} ({x \over y}) \otimes g_-(y)),  \nonumber \\
{dg_+(x) \over{dt}}=
{\alpha_s \over {2 \pi}} (P_{++}^{gq} ({x \over y}) \otimes q_+(y)+
P_{+-}^{gq} ({x \over y}) \otimes q_-(y) \nonumber \\
+P_{++}^{gg} ({x \over y}) \otimes g_+(y)+
P_{+-}^{gg} ({x \over y}) \otimes g_-(y)),  \nonumber \\
{dg_-(x) \over{dt}}=
{\alpha_s \over {2 \pi}} (P_{+-}^{gq} ({x \over y}) \otimes q_+(y)+
P_{++}^{gq} ({x \over y}) \otimes q_-(y) \nonumber \\
+P_{+-}^{g} ({x \over y}) \otimes g_+(y)+
P_{++}^{gg} ({x \over y}) \otimes g_-(y)).
\label{hs}\end{eqnarray}
The non-singlet (valence) analogue of this equation is also easy to
write down
\begin{eqnarray}
{dq_{+, V}(x) \over{dt}}=
{\alpha_s \over {2 \pi}} (P_{++} ({x \over y}) \otimes q_{+,V}(y)+
P_{+-} ({x \over y}) \otimes q_{-,V}(y)), \nonumber \\
{dq_{-,V}(x) \over{dt}}=
{\alpha_s \over {2 \pi}} (P_{+-} ({x \over y}) \otimes q_{+,V}(y)+
P_{++} ({x \over y}) \otimes q_{-,V}(y)).
\label{h}\end{eqnarray}
where the $q_{\pm,V}=q_\pm - \bar{q}_\pm$ are the valence components of fixed helicity.
The kernels in this basis are given by 
\beqa
P_{NS\pm,++}^{(0)} &=&P_{qq, ++}^{(0)}=P_{qq}^{(0)}\nonumber \\
P_{qq,+-}^{(0)}&=&P_{qq,-+}^{(0)}= 0\nonumber \\
P_{qg,++}^{(0)}&=& n_f x^2\nonumber \\
P_{qg,+-}&=& P_{qg,-+}= n_f(x-1)^2 \nonumber \\
P_{gq,++}&=& P_{gq,--}=C_F\frac{1}{x}\nonumber \\ 
P_{gg,++}^{(0)}&=&P_{gg,++}^{(0)}= N_c
\left(\frac{2}{(1-x)_+} +\frac{1}{x} -1 -x - x^2 \right) +{\beta_0}\delta(1-x) \nonumber \\
P_{gg,+-}^{(0)}&=& N_c
\left( 3 x +\frac{1}{x} -3 - x^2 \right). 
\label{stand1}
\eeqa

Taking linear combinations of these equations (adding and subtracting),
one recovers the usual evolutions for unpolarized $q(x)$ and longitudinally
polarized $\Delta q(x)$ distributions.

\section{The Ansatz and some Examples}

In order to solve the evolution equations directly in \( x \)-space,
we assume solutions of the form\begin{equation}
\label{eq:ansatz}
f(x,Q^{2})=\sum _{n=0}^{\infty }\frac{A_{n}(x)}{n!}\log ^{n}\frac{\alpha _{s}(Q^{2})}{\alpha _{s}(Q_{0}^{2})}+\alpha _{s}(Q^{2})\sum _{n=0}^{\infty }\frac{B_{n}(x)}{n!}\log ^{n}\frac{\alpha _{s}(Q^{2})}{\alpha _{s}(Q_{0}^{2})},
\label{rossis}
\end{equation}
for each parton distribution \( f \), where $Q_0$ defines the initial 
evolution scale. The justification of this ansatz can be found, 
at least in the case of the photon structure function, 
in the original work of Rossi \cite{Rossi}, and its connection 
to the ordinary solutions of the DGLAP equations is most easily 
worked out by taking moments of the scale invariant coefficient 
functions $A_n$ and $B_n$ and comparing them to 
the corresponding moments 
of the parton distributions, as we are going to illustrate 
in section 5. The link between Rossi's expansion 
and the solution of the evolution equations 
(which are ordinary differential equations) in the space 
of the moments up to NLO will be discussed in that section, from which 
it will be clear that 
Rossi's ansatz involves a resummation 
of the ordinary Mellin moments of the parton distributions.   

Setting \( Q=Q_{0} \) in (\ref{eq:ansatz})
we get\begin{equation}
\label{eq:boundary}
f(x,Q_{0}^{2})=A_{0}(x)+\alpha _{s}(Q_{0}^{2})B_{0}(x).
\end{equation}
Inserting (\ref{eq:ansatz}) in the evolution equations, we obtain
the following recursion relations for the coefficients \( A_{n} \)
and \( B_{n} \) 
\begin{equation}
\label{eq:An_recurrence}
A_{n+1}(x)=-\frac{2}{\beta _{0}}P^{(0)}(x)\otimes A_{n}(x),
\end{equation}
\begin{equation}
\label{eq:Bn_recurrence}
B_{n+1}(x)=-B_{n}(x)-\frac{\beta _{1}}{4\beta _{0}}A_{n+1}(x)-\frac{2}{\beta _{0}}P^{(0)}(x)\otimes B_{n}(x)-\frac{1}{4\pi \beta _{0}}P^{(1)}(x)\otimes A_{n}(x), 
\end{equation}
obtained by equating left-hand sides and right-hand-side of the equation 
of the same logarithmic power 
in $\log^n\alpha_s(Q^2)$ and $\alpha_s \log^n \alpha_s(Q^2)$.

Any boundary condition satisfying (\ref{eq:boundary}) can be chosen at the lowest scale $Q_0$ and in our case we choose\begin{equation}
\label{eq:initial}
B_{0}(x)=0,\qquad f(x,Q_{0}^{2})=A_{0}(x).
\end{equation}

The actual implementation of the recursion relations is the main effort 
in the actual writing of the code. Obviously, 
this requires particular care in the handling of 
the singularities in $x-$space, being all the kernels defined as 
distributions. Since the distributions are integrated, 
there are various ways to render the integrals finite, 
as discussed in the previous literature on the method \cite{Storrow} 
in the case of the photon structure function. 
In these previous studies the 
edge-point contributions - i.e. the terms which multiply $\delta(1-x)$ 
in the kernels - are approximated using a sequence of functions 
converging to the $\delta$ function in a distributional sense.

This technique is not very efficient. We think that 
the best way to proceed is to actually perform the integrals explicitly in the recursion relations and let the subtracting 
terms appear under the same integral together with the 
bulk contributions ($x<1$) (see also \cite{Gordon}). This procedure is best exemplified 
by the integral relation 
\beq
\int_x^1 \frac{dy}{y (1-y)_+}f(x/y)=\int_x^1\frac{dy}{y}
\frac{ yf(y) - x f(x)}{y-x} -\log(1-x) f(x)
\label{simplerel}
\eeq
in which, on the right hand side, regularity of both the first 
and the second term is explicit. For instance, the evolution equations become 
(prior to separation between singlet and non-singlet sectors) in the unpolarized 
case
\beqa
\frac{d q_i(x)}{d \log(Q^2)} &=& 2 C_F 
\int\frac{dy}{y}\frac{ y q_i(y) - x q_i(x)}{y-x} +2 C_F \log(1-x) q_i(x) -
\int_x^1\frac{dy}{y}\left( 1 + z\right)q_i(y) + 
\frac{3}{2} C_F q(x) \nonumber \\
&& + n_f\int_x^1\frac{dy}{y}
\left( z^2 +(1-z)^2\right)g(y)\nonumber \\
\frac{d g(x)}{d \log(Q^2)} &=& 
C_F \int_x^1\frac{dy}{y}\frac{1 +(1-z)^2}{z}q_i(y)
+ 2 N_c \int_x^1\frac{dy}{y}
\frac{ y f(y) - x f(x)}{y-x}g(y) 
\nonumber \\
&& + 2 N_c \log(1-x) g(x) 
+2 N_c\int_x^1 \frac{dy}{y}\left( \frac{1}{z} -2 + z(1-z)\right)g(y) + 
\frac{\beta_0}{2}g(x) \nonumber \\
\label{standard}
\eeqa
with $z\equiv x/y$. Here $q$ are fixed flavour distributions.

\section{An Example: The Evolution of the Transverse Spin Distributions}
Leading order (LO) and NLO recursion relations for the coefficients of the expansion 
can be worked out quite easily. We illustrate here in detail 
the implementation of a non-singlet evolutions, such 
as those involving transverse spin distributions. 
For the first recursion relation (\ref{eq:An_recurrence}) in this case 
we have
\beqn
&&A^{\pm}_{n+1}(x)=-\frac{2}{\beta_{0}}\Delta_{T}P^{(0)}_{qq}(x)\otimes A^{\pm}_{n}(x)=\nonumber\\ 
&&C_{F}\left(-\frac{4}{\beta_{0}}\right)\left[\int^{1}_{x}\frac{dy}{y}\frac{y A^{\pm}_{n}(y) - x A^{\pm}_{n}(x)}{y-x} + A^{\pm}_{n}(x) \log(1-x)\right]+\nonumber\\
&&C_{F}\left(\frac{4}{\beta_{0}}\right) \left(\int_{x}^{1}\frac{dy}{y} A^{\pm}_{n}(y)\right) + C_{F}\left(-\frac{2}{\beta_{0}}\right)\frac{3}{2} A^{\pm}_{n}(x)\,.
\eeqn
As we move to NLO, it is convenient to summarize 
the structure of the transverse kernel $\Delta_{T}P^{\pm, (1)}_{qq}(x)$ as  

\beqn
&&\Delta_{T}P^{\pm, (1)}_{qq}(x)= K^{\pm}_{1}(x)\delta(1-x) + K^{\pm}_{2}(x)S_{2}(x) +K^{\pm}_{3}(x)\log(x)\nonumber\\
&&+ K^{\pm}_{4}(x)\log^{2}(x) +K^{\pm}_{5}(x)\log(x)\log(1-x) + K^{\pm}_{6}(x)\frac{1}{(1-x)_{+}} + K^{\pm}_{7}(x)\,.    
\eeqn

Hence, for the $(+)$ case we have 

\beqn
&&\Delta_{T}P^{+, (1)}_{qq}(x)\otimes A^{+}_{n}(x) = K^{+}_{1} A^{+}_{n}(x) + \int^{1}_{x}\frac{dy}{y}\left[K^{+}_{2}(z) S_{2}(z) + K^{+}_{3}(z)\log(z) \right.\nonumber\\
&& \left. + \log^{2}(z)K^{+}_{4}(z) + \log(z)\log(1-z)K^{+}_{5}(z)\right] A^{+}_{n}(y) +  \nonumber\\ 
&&K^{+}_{6}\left\{\int^{1}_{x}\frac{dy}{y} \frac{yA^{+}_{n}(y) - xA^{+}_{n}(x)}{y-x} + A^{+}_{n}(x)\log(1-x) \right\} + K^{+}_{7}\int^{1}_{x}\frac{dy}{y}A^{+}_{n}(y)\,, 
\eeqn

where $z={x}/{y}$. For the $(-)$ case we get a similar expression.
  
For the $B^{\pm}_{n+1}(x)$  we get (for the $(+)$ case) 

\ba
&&B^{+}_{n+1}(x) = - B^{+}_{n}(x) + \frac{\beta_{1}}{2\beta^{2}_{0}} \left\{2C_{F}\left[\int^{1}_{x}\frac{dy}{y}\frac{y A^{+}_{n}(y) - x A^{+}_{n}(x)}{y-x} + A^{+}_{n}(x) \log(1-x)\right]\right.+\nonumber\\
&&\left.-2C_{F}\left(\int_{x}^{1}\frac{dy}{y} A^{+}_{n}(y)\right) + C_{F}\frac{3}{2} A^{+}_{n}(x)\right\}-\frac{1}{4\pi\beta_{0}}K^{+}_{1} A^{+}_{n}(x)+ \int^{1}_{x}\frac{dy}{y}\left[ K^{+}_{2}(z) S_{2}(z) + \right.\nonumber\\
&&+ \left.K^{+}_{3}(z)\log(z)+\log^{2}(z)K^{+}_{4}(z) + \log(z)\log(1-z)K^{+}_{5}(z)\right]\left(-\frac{1}{4\pi\beta_{0}}\right)A^{+}_{n}(y)+\nonumber\\
&&K^{+}_{6}\left(-\frac{1}{4\pi\beta_{0}}\right)\left\{\left[\int^{1}_{x}\frac{dy}{y} \frac{yA^{+}_{n}(y) - xA^{+}_{n}(x)}{y-x} + A^{+}_{n}(x)\log(1-x) \right] + K^{+}_{7}\int^{1}_{x}\frac{dy}{y}A^{+}_{n}(y)\right\}-\nonumber\\
&&C_{F}\left(-\frac{4}{\beta_{0}}\right)\left[\int^{1}_{x}\frac{dy}{y}\frac{y B^{\pm}_{n}(y) - x B^{\pm}_{n}(x)}{y-x} + B^{\pm}_{n}(x) \log(1-x)\right]+\nonumber\\
&&C_{F}\left(\frac{4}{\beta_{0}}\right) \left(\int_{x}^{1}\frac{dy}{y} B^{\pm}_{n}(y)\right) + C_{F}\left(-\frac{2}{\beta_{0}}\right)\frac{3}{2} B^{\pm}_{n}(x)\,\nonumber
\ea
where in the $(+)$ case we have the expressions 

\ba
&&K_{1}^{+}(x)=\frac{1}{72}C_{F} (-2 n_{f} (3+4\pi^{2}) + N_{C}(51 + 44\pi^{2} - 216 \zeta(3))+ 9C_{F}(3-4\pi^{2}+48\zeta(3))\nonumber\\
&&K_{2}^{+}(x)= \frac{2 C_{F}(-2C_{F}+N_{C})x}{1+x}\nonumber \\
&&K_{3}^{+}(x)= \frac{C_{F}(9C_{F}-11 N_{C}+2n_{f})x}{3(x-1)}\nonumber \\
&&K_{4}^{+}(x)=\frac{C_{F}N_{C}x}{1-x}\nonumber \\ 
&&K_{5}^{+}(x)=\frac{4C_{F}^{2}x}{1-x}\nonumber \\
&&K_{6}^{+}(x)=-\frac{1}{9}C_{F}(10n_{f}+N_{C}(-67 + 3\pi^{2}))\nonumber\\
&&K_{7}^{+}(x)=\frac{1}{9}C_{F}(10n_{f}+N_{C}(-67 + 3\pi^{2})),\nonumber\\
\ea
and for the $(-)$ case
\ba
&&K_{1}^{-}(x)=\frac{1}{72}C_{F} (-2 n_{f} (3+4\pi^{2}) + N_{C}(51 + 44\pi^{2} - 216 \zeta(3))+ 9C_{F}(3-4\pi^{2}+48\zeta(3))\nonumber\\
&&K_{2}^{-}(x)= \frac{2 C_{F}(+2C_{F}-N_{C})x}{1+x}\nonumber \\
&&K_{3}^{-}(x)= \frac{C_{F}(9C_{F}-11 N_{C}+2n_{f})x}{3(x-1)}\nonumber\\
&&K_{4}^{-}(x)=\frac{C_{F}N_{C}x}{1-x}\nonumber\\
&&K_{5}^{-}(x)=\frac{4C_{F}^{2}x}{1-x}\nonumber\\
&&K_{6}^{-}(x)=-\frac{1}{9}C_{F}(10n_{f}+N_{C}(-67 + 3\pi^{2}))\nonumber\\
&&K_{7}^{-}(x)=-\frac{1}{9}C_{F}(10n_{f}-18 C_{F}(x-1)+N_{C}
(-76 +3\pi^{2}+9x)).\nonumber\\
\ea
The terms containing similar distribution (such as ``+'' distributions 
and $\delta$ functions) have been combined 
together in order to speed-up the computation of the recursion relations. 

\section{Comparisons among Moments}

It is particularly instructing to illustrate here briefly the relation 
between the Mellin moments of the parton distributions, which evolve 
with standard ordinary differential equations, and those of the 
arbitrary coefficient $A_n(x)$ and $B_n(x)$ which characterize 
Rossi's expansion up to next-to leading order (NLO). This relation, as we are going to show, involves a 
resummation of the ordinary moments of the parton distributions. 

Specifically, here we will be dealing with the relation between 
the Mellin moments of the coefficients appearing in the expansion 
\beqa
A(N) &=& \int_0^1\,dx \, x^{N-1} A(x)\nonumber \\
B(N) &=&\int_0^1\,dx \, x^{N-1} B(x) \nonumber \\
\eeqa
and those of the distributions  
\beq
\Delta_T q^{(\pm)}(N,Q^2)=\int_0^1\,dx \, x^{N-1} \Delta_T q^{(\pm)}(x,Q^2)).
\eeq 
For this purpose we recall that the general (non-singlet) solution to NLO for the latter moments is given by 
\begin{eqnarray} \label{evsol}
\nonumber
\Delta_T q_{\pm} (N,Q^2) &=& K(Q_0^2,Q^2,N)
\left( \frac{\alpha_s (Q^2)}{\alpha_s (Q_0^2)}\right)^{-2\Delta_T 
P_{qq}^{(0)}(N)/ \beta_0}\! \Delta_T q_{\pm}(N, Q_0^2)
\label{solution}
\end{eqnarray}
with the input distributions $\Delta_T q_{\pm}^n (Q_0^2)$ at the input scale 
$Q_0$.
We also have set 
\beq
K(Q_0^2,Q^2,N)= 1+\frac{\alpha_s (Q_0^2)-
\alpha_s (Q^2)}{\pi\beta_0}\!
\left[ \Delta_T P_{qq,\pm}^{(1)}(N)-\frac{\beta_1}{2\beta_0} \Delta_T 
P_{qq\pm}^{(0)}(N) \right]. 
\eeq
In the expressions above we have introduced the corresponding moments for the LO and NLO kernels 
($\Delta_T P_{qq}^{(0),N}$,
$ \Delta_T P_{qq,\pm}^{(1),N})$. 

The relation between the moments of the coefficients of the non-singlet
$x-$space expansion and those of the parton distributions at any $Q$, as expressed by eq.~(\ref{solution}) can be easily written down
\beq
A_n(N) + \alpha_s B_n(N)=\Delta_T q_\pm(N,Q_0^2)K(Q_0,Q,N)\left(\frac{-2 \Delta_T P_{qq}(N)}{\beta_0}\right)^n.
\label{relation}
\eeq

As a check of this expression, notice that the initial condition is easily obtained from  
(\ref{relation}) setting $Q\to Q_0, n\to 0$, thereby obtaining 
\beq
A_0^{NS}(N) + \alpha_s B_0^{NS} (N)= \Delta_T q_\pm(N,Q_0^2),
\eeq
which can be solved with $A_0^{NS}(N)=\Delta_T q_\pm(N,Q_0^2)$ and 
$B_0^{NS} (N)=0$. 

It is then evident that the expansion (\ref{rossis}) involves a resummation of the logarithmic contributions, as shown in eq.~(\ref{relation}). 

In the singlet sector we can work out a similar relation both to LO

\beq
A_n(N) = e_1\left(\frac{-2 \lambda_1}{\beta_0}\right)^n 
+e_2 \left(\frac{-2 \lambda_2}{\beta_0}\right)^n 
\eeq

with 
\beqa
e_1 &=& \frac{1}{\lambda_1 - \lambda_2}\left( P^{(0)}(N)- \lambda_2 \one \right)
\nonumber \\
e_2 &=& \frac{1}{\lambda_2 - \lambda_1}\left( - P^{(0)}(N) + \lambda_1 \one\right)
\nonumber \\
\lambda_{1,2}&=& \frac{1}{2}\left( 
P^{(0)}_{qq}(N) + P^{(0)}_{gg}(N) \pm \sqrt{\left(P^{(0)}_{qq}(N)- P^{(0)}_{gg}(N)\right)^2  
+ 4 P^{(0)}_{qg}(N)P^{(0)}_{gq}(N)}\right),
\eeqa
and to NLO 
\beq
A_n(N) + \alpha_s B_n(N) = \chi_1\left(\frac{-2 \lambda_1}{\beta_0}\right)^n 
+\chi_2 \left(\frac{-2 \lambda_2}{\beta_0}\right)^n, 
\eeq

where 
\beqa
\chi_1 &=& e_1 + \frac{\alpha}{2 \pi}\left( \frac{-2}{\beta_0}e_1 \R e_1 
+\frac{ e_2 \R e_1}{\lambda_1 - \lambda_2 - \beta_0/2}\right) \nonumber \\
\chi_2 &=& e_2 + \frac{\alpha}{2 \pi}\left( \frac{-2}{\beta_0}e_2 \R e_2  
+\frac{ e_1 \R e_2}{\lambda_2 - \lambda_1 - \beta_0/2}\right)\nonumber \\
\eeqa
with 
\beq
\R= P^{(1)}(N) -\frac{\beta_1}{2 \beta_0}P^{(0)}(N).
\eeq
We remark, if not obvious, that $A_n(N)$ and $B_n(N)$, $P^{(0)}(N)$, $P^{(1)}(N)$, in this case, 
are all 2-by-2 singlet matrices.

\section{Initial conditions\label{sec:initial}}

As input distributions in the unpolarized case, we have used the 
models of Ref.\cite{GRV}, valid to NLO in the \( \overline{\textrm{MS}} \)
scheme at a scale \( Q_{0}^{2}=0.40\, \textrm{GeV}^{2} \)
\begin{eqnarray}
x(u-\overline{u})(x,Q_{0}^{2}) & = & 0.632x^{0.43}(1-x)^{3.09}(1+18.2x)\nonumber \\
x(d-\overline{d})(x,Q_{0}^{2}) & = & 0.624(1-x)^{1.0}x(u-\overline{u})(x,Q_{0}^{2})\nonumber \\
x(\overline{d}-\overline{u})(x,Q_{0}^{2}) & = & 0.20x^{0.43}(1-x)^{12.4}(1-13.3\sqrt{x}+60.0x)\nonumber \\
x(\overline{u}+\overline{d})(x,Q_{0}^{2}) & = & 1.24x^{0.20}(1-x)^{8.5}(1-2.3\sqrt{x}+5.7x)\nonumber \\
xg(x,Q_{0}^{2}) & = & 20.80x^{1.6}(1-x)^{4.1}
\end{eqnarray}
and \( xq_{i}(x,Q_{0}^{2})=x\overline{q_{i}}(x,Q_{0}^{2})=0 \) for
\( q_{i}=s,c,b,t \).

Following \cite{GRSV}, we have related the unpolarized input distribution
to the longitudinally polarized ones 

\begin{eqnarray}
x\Delta u(x,Q_{0}^{2}) & = & 1.019x^{0.52}(1-x)^{0.12}xu(x,Q_{0}^{2})\nonumber \\
x\Delta d(x,Q_{0}^{2}) & = & -0.669x^{0.43}xd(x,Q_{0}^{2})\nonumber \\
x\Delta \overline{u}(x,Q_{0}^{2}) & = & -0.272x^{0.38}x\overline{u}(x,Q_{0}^{2})\nonumber \\
x\Delta \overline{d}(x,Q_{0}^{2}) & = & x\Delta \overline{u}(x,Q_{0}^{2})\nonumber \\
x\Delta g(x,Q_{0}^{2}) & = & 1.419x^{1.43}(1-x)^{0.15}xg(x,Q_{0}^{2})
\end{eqnarray}
and \( x\Delta q_{i}(x,Q_{0}^{2})=x\Delta \overline{q_{i}}(x,Q_{0}^{2})=0 \)
for \( q_{i}=s,c,b,t \).
Being the transversity
distribution experimentally unknown, following \cite{MSSV}, 
we assume the saturation of
Soffer's inequality
\begin{equation}
x\Delta _{T}q_{i}(x,Q_{0}^{2})=\frac{xq_{i}(x,Q_{0}^{2})+x\Delta q_{i}(x,Q_{0}^{2})}{2}.
\end{equation}

\section{Names of the input parameters, variables and of the output files}

\subsection{Notations \label{subsec:dist_ind}}

\begin{tabular}{lll}
\hline 
0&
gluons, \( g \)&
\texttt{g}\\
1-6&
quarks, \( q_{i} \), sorted by their mass values(\( u,d,s,c,b,t \))&
\texttt{u,d,s,c,b,t}\\
7-12&
antiquarks, \( \overline{q_{i}} \)&
\texttt{au,ad,as,ac,ab,at}\\
13-18&
\( q^{(-)}_{i} \)&
\texttt{um,dm,sm,cm,bm,tm}\\
19-24&
\( \chi _{i} \) (unpolarized and longitudinally polarized cases)&
\texttt{Cu,Cd,Cs,Cc,Cb,Ct}\\
&
\( q^{(+)}_{i} \) (transversely polarized case)&
\texttt{Cu,Cd,Cs,Cc,Cb,Ct}\\
25&
\( q^{(+)} \)&
\texttt{qp}\\
\hline
\end{tabular}

\subsection{Input parameters and variables}

\begin{tabular}{lll}
\hline 
\texttt{process}&
0&
unpolarized\\
&
1&
longitudinally polarized\\
&
2&
transversely polarized\\
\hline
\end{tabular}\\
\begin{tabular}{lll}
\hline 
\texttt{spacing}&
1&
linear\\
&
2&
logarithmic\\
\end{tabular}\\
\begin{tabular}{ll}
\hline 
\texttt{GRID\_PTS}&
Number of points in the grid\\
\texttt{NGP}&
Number of Gaussian points, \( n_{G} \)\\
\texttt{ITERATIONS}&
Number of terms in the sum (\ref{eq:ansatz})\\
\texttt{extension}&
Extension of the output files\\
\hline
\end{tabular}\\
\begin{tabular}{ll}
\hline 
\texttt{step}&
grid step (linear spacing case)\\
\texttt{lstep}&
step in \( \log _{10}x \) (logarithmic spacing case)\\
\texttt{X{[}i{]}}&
\( i \)-th grid point, \( x_{i} \)\\
\texttt{XG{[}i{]}{[}j{]}}&
\( j \)-th Gaussian abscissa in the range \( [x_{i},1] \), \( X_{ij} \)\\
\texttt{WG{[}i{]}{[}j{]}}&
\( j \)-th Gaussian weights in the range \( [x_{i},1] \), \( W_{ij} \)\\
\texttt{nf, Nf}&
number of active flavors, \( n_{f} \)\\
\texttt{n\_evol}&
progressive number of the evolution step is \( n_{f}-3 \)\\
\texttt{Q{[}i{]}}&
values of \( Q \)  in the corresponding grid \\
\texttt{lambda{[}i{]}}&
\( \Lambda _{\overline{MS}}^{(n_{f})} \), where \( i=n_{f}-3 \)\\
\texttt{A{[}i{]}{[}j{]}{[}k{]}}&
coefficient \( A_{j}(x_{k}) \) for the distribution with index \( i \)\\
\texttt{B{[}i{]}{[}j{]}{[}k{]}}&
coefficient \( B_{j}(x_{k}) \) for the distribution with index \( i \)\\
\texttt{beta0}&
\( \beta _{0} \)\\
\texttt{beta1}&
\( \beta _{1} \)\\
\texttt{alpha1}&
\( \alpha _{s}(Q_{in}) \), where \( Q_{in} \) is the lower \( Q \)
of the evolution step\\
\texttt{alpha2}&
\( \alpha _{s}(Q_{fin}) \), where \( Q_{fin} \) is the higher \( Q \)
of the evolution step\\
\hline
\end{tabular}

\subsection{Output files}

The generic name of an output file is: \texttt{\textbf{X}}\texttt{\textbf{\emph{n}}}\texttt{\textbf{i.ext}},
where

\begin{description}
\item [\texttt{X}]is U in the unpolarized case, L in the longitudinally
polarized case and T in the transversely polarized case;
\item [\texttt{\emph{n}}]is a progressive number that indicates the scale
\( Q^{2} \) at a given stage: \( n=0 \) refers to the initial
scale, the highest value of \( n \) refers to the final scale and
the intermediate values refer to the quarks production thresholds
(1 for charm, 2 for bottom and 3 for top);
\item [\texttt{i}]is the identifier of the distribution, reported in
the third column of the table in subsection \ref{subsec:dist_ind};
\item [\texttt{ext}]is an extension chosen by the user.
\end{description}

\section{Description of the program}

\subsection{Main program}

At run time, the program asks the user to select a linear or a logarithmic
spacing for the \( x \)-axis. The logarithmic spacing is useful in order 
to analyze the small-\( x \) behavior. Then the program stores as external
variables the grid points \( x_{i} \) and, for each of them, calls
the function \texttt{gauleg} which computes the Gaussian points
\( X_{ij} \) and weights \( W_{ij} \) corresponding to the integration
range \( [x_{i},1] \), with \( 0\leq j\leq n_{G}-1 \). After that,
the user is asked to enter the type of process, the final value
of \( Q \) and an extension for the names of the output files. At this point the
program computes the initial values of the parton distributions for
gluons, up, down, anti-up and anti-down (see Section \ref{sec:initial})
at the grid points and stores them in the arrays \texttt{A{[}i{]}{[}0{]}{[}k{]}}
(see (\ref{eq:initial})), setting to zero the initial distributions
of the heavier quarks.

The evolution is done in the various regions of the evolutions, all characterized by a specific flavour number. 
Each new flavour comes into play
only when the scale $Q$ reaches the corresponding quark mass. 
In that case $n_f$ is increased by 1 everywhere in the program.
The recurrence relations (\ref{eq:An_recurrence}) and (\ref{eq:Bn_recurrence})
are then solved iteratively for both the nonsinglet and the singlet
sector, and at the end of each energy step the evolved distributions
are reconstructed via the relation (\ref{eq:ansatz}). The distributions
computed in this way become the initial conditions for the subsequent
step. The numerical values of the distributions at the end of each
energy step are printed to files.

\subsection{Function \texttt{writefile}}

\texttt{void writefile(double {*}A,char {*}filename);}

This function creates a file, whose name is contained in the string
\texttt{{*}filename}, with an output characterized by 
two columns of data: the left column contains
all the values of the grid points \( x_{i} \) 
and the right one the corresponding values of the array \texttt{A{[}i{]}}.

\subsection{Function \texttt{alpha\_s}}

\texttt{double alpha\_s(double Q,double beta0,double beta1,double
lambda);}

Given the energy scale \texttt{Q}, the first two terms of the perturbative
expansion of the \( \beta  \)-function \texttt{beta0} and \texttt{beta1}
and the value \texttt{lambda} of \( \Lambda _{\overline{MS}}^{(n_{f})} \),
\texttt{alpha\_s} returns the two-loop running of the coupling constant,
using the formula (\ref{eq:alpha_s}).

\subsection{Function \texttt{gauleg}}

\texttt{void gauleg(double x1,double x2,double x{[}{]},double w{[}{]},int
n);}

This function is taken from \cite{NumRecipes} with just some minor changes.
Given the lower and upper limits of integration \texttt{x1} and \texttt{x2},
and given \texttt{n}, \texttt{gauleg} returns arrays \texttt{x{[}0,...,n-1{]}}
and \texttt{w{[}0,...,n-1{]}} of length \texttt{n}, containing the
abscissas and weights of the Gauss-Legendre \texttt{n}-point quadrature
formula.

\subsection{Function \texttt{interp}}

\texttt{double interp(double {*}A,double x);}

Given an array \texttt{A}, representing a function known only at the
grid points, and a number \texttt{x} in the interval \( [0,1] \),
\texttt{interp}
returns the linear interpolation of the function at
the point \texttt{x}.

\subsection{Function \texttt{IntGL}}

\texttt{double IntGL(int i,double kernel(double z),double {*}A);}

Given an integer \texttt{i} (corresponding to a grid point \( x_{i} \)),
a one variable function \texttt{kernel(z)} and an array \texttt{A},
representing a function \( g(x) \) known at the grid points, \texttt{IntGL}
returns the result of the integral
\begin{equation}
\int _{x_{i}}^{1}\frac{\textrm{d}y}{y}kernel\left( \frac{x_{i}}{y}\right) g(x_{i}),
\end{equation}
computed by the Gauss-Legendre technique.

\subsection{Function \texttt{IntPD}}

\texttt{double IntGL(int i,double {*}A);}

Given an integer \texttt{i}, to which it corresponds a grid point \( x_{i} \),
and an array \texttt{A}, representing a function \( f(x) \) known
at the grid points, \texttt{IntGL} returns the result of the convolution\begin{equation}
\frac{1}{(1-x_{i})_{+}}\otimes f(x_{i})=\int _{x_{i}}^{1}\frac{\textrm{d}y}{y}\frac{yf(y)-x_{i}f(x_{i})}{y-x_{i}}+f(x_{i})\log (1-x_{i}),
\end{equation}
computed by the Gauss-Legendre technique.

\subsection{Function \texttt{S2}}

\texttt{double S2(double z);}

This function evaluates the Spence function
\( S_{2}(z) \) using the expansion\begin{equation}
S_{2}(z)=\log z\log (1-z)-\frac{1}{4}\left( \log z\right) ^{2}+\frac{\pi ^{2}}{12}+\sum ^{\infty }_{n=1}\frac{(-z)^{n}}{n^{2}}
\end{equation}
arrested at the 50th order.

\subsection{Function \texttt{fact}}

\texttt{double fact(int n);}

This function returns the factorial \( n! \)

\subsection{Initial distributions}

\texttt{double xuv(double x);}~\\
\texttt{double xdv(double x);}~\\
\texttt{double xdbmub(double x);}~\\
\texttt{double xubpdb(double x);}~\\
\texttt{double xg(double x);}~\\
\texttt{double xu(double x);}~\\
\texttt{double xubar(double x);}~\\
\texttt{double xd(double x);}~\\
\texttt{double xdbar(double x);}~\\
\texttt{double xDg(double x);}~\\
\texttt{double xDu(double x);}~\\
\texttt{double xDubar(double x);}~\\
\texttt{double xDd(double x);}~\\
\texttt{double xDdbar(double x);}

Given the Bjorken variable \texttt{x}, these functions return the
initial distributions at the input scale (see Section \ref{sec:initial}).

\subsection{Regular part of the kernels}

\texttt{double P0NS(double z);}~\\
\texttt{double P0qq(double z);}~\\
\texttt{double P0qg(double z);}~\\
\texttt{double P0gq(double z);}~\\
\texttt{double P0gg(double z);}~\\
\texttt{double P1NSm(double z);}~\\
\texttt{double P1NSp(double z);}~\\
\texttt{double P1qq(double z);}~\\
\texttt{double P1qg(double z);}~\\
\texttt{double P1gq(double z);}~\\
\texttt{double P1gg(double z);}~\\
\texttt{double DP0NS(double z);}~\\
\texttt{double DP0qq(double z);}~\\
\texttt{double DP0qg(double z);}~\\
\texttt{double DP0gq(double z);}~\\
\texttt{double DP0gg(double z);}~\\
\texttt{double DP1NSm(double z);}~\\
\texttt{double DP1NSp(double z);}~\\
\texttt{double DP1qq(double z);}~\\
\texttt{double DP1qg(double z);}~\\
\texttt{double DP1gq(double z);}~\\
\texttt{double DP1gg(double z);}~\\
\texttt{double tP0(double z);}~\\
\texttt{double tP1m(double z);}~\\
\texttt{double tP1p(double z);}

Given the Bjorken variable \texttt{z}, these functions return the
part of the Altarelli-Parisi kernels that does not contain singularities.

\section{Running the code}

In the plots shown in this paper we have divided the interval \( [0,1] \)
of the Bjorken variable \( x \) in 500 subintervals (\texttt{GRID\_PTS}=501),
30 Gaussian points (\texttt{NGP}=1), and we have retained 10 terms
in the sum (\ref{eq:ansatz}) (\texttt{ITERATIONS}=10). In the figures
\ref{fig:Lu_log} and \ref{fig:Lg_log} the flag \texttt{spacing}
has been set to 2, in order to have a logarithmically spaced grid. This feature
turns useful if one intends to analyze the small-\( x \) behavior. We have tested our 
implementation in a detailed study of Soffer's inequality up to NLO 
\cite{CafarellaCorianoGuzzi}.

\begin{figure}[tbh]
{\centering \resizebox*{8cm}{!}{\rotatebox{-90}{\includegraphics{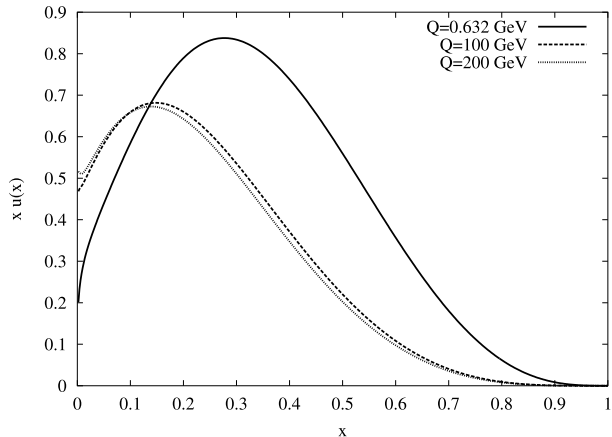}}} \par}

\caption{Evolution of the unpolarized quark up distribution \protect\( xu\protect \)
versus \protect\( x\protect \) at various \protect\( Q\protect \)
values.}
\end{figure}

\begin{figure}[tbh]
{\centering \resizebox*{8cm}{!}{\rotatebox{-90}{\includegraphics{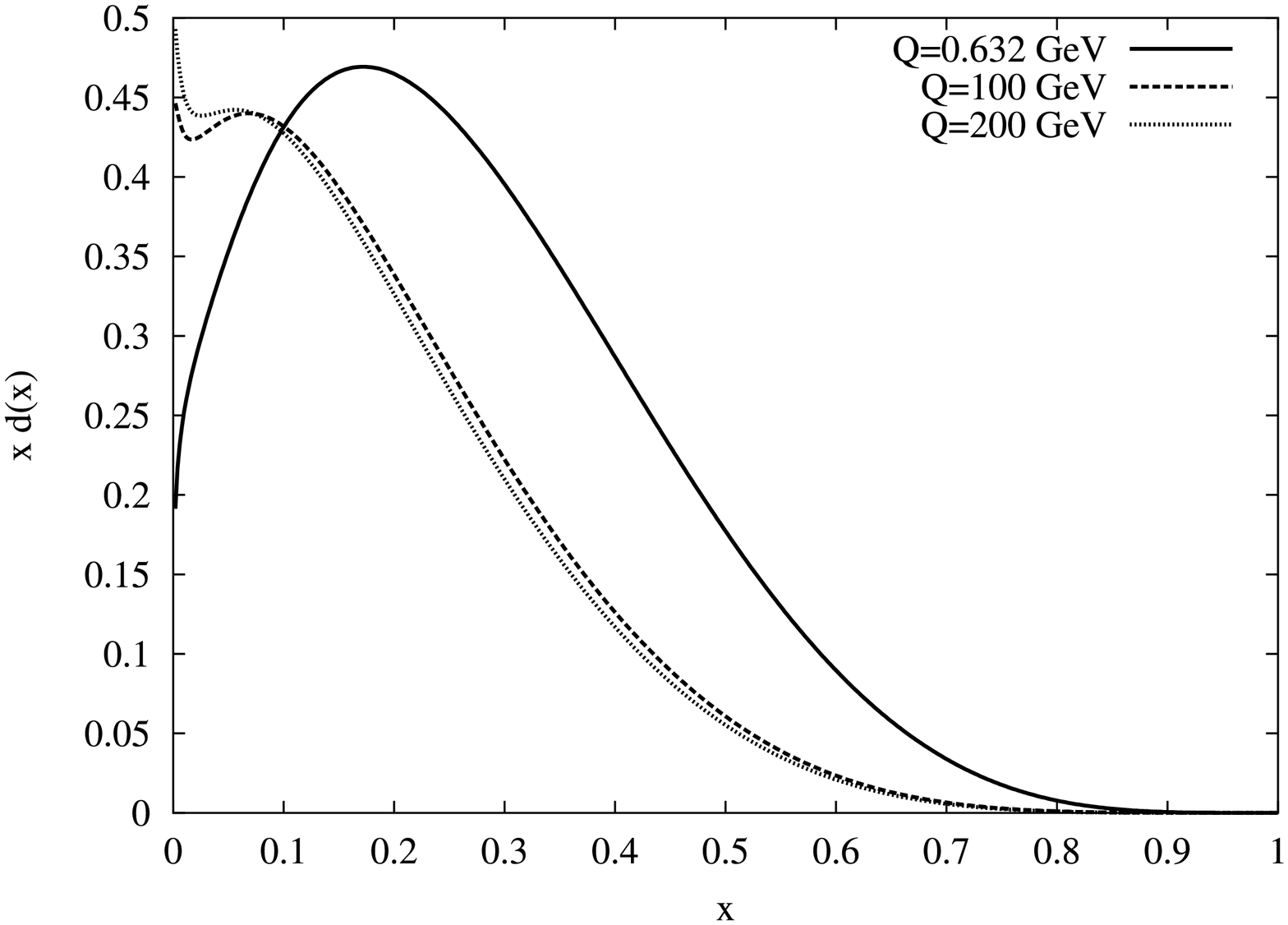}}} \par}

\caption{Evolution of \protect\( xd\protect \) versus \protect\( x\protect \)
at various \protect\( Q\protect \) values.}
\end{figure}

\begin{figure}[tbh]
{\centering \resizebox*{8cm}{!}{\rotatebox{-90}{\includegraphics{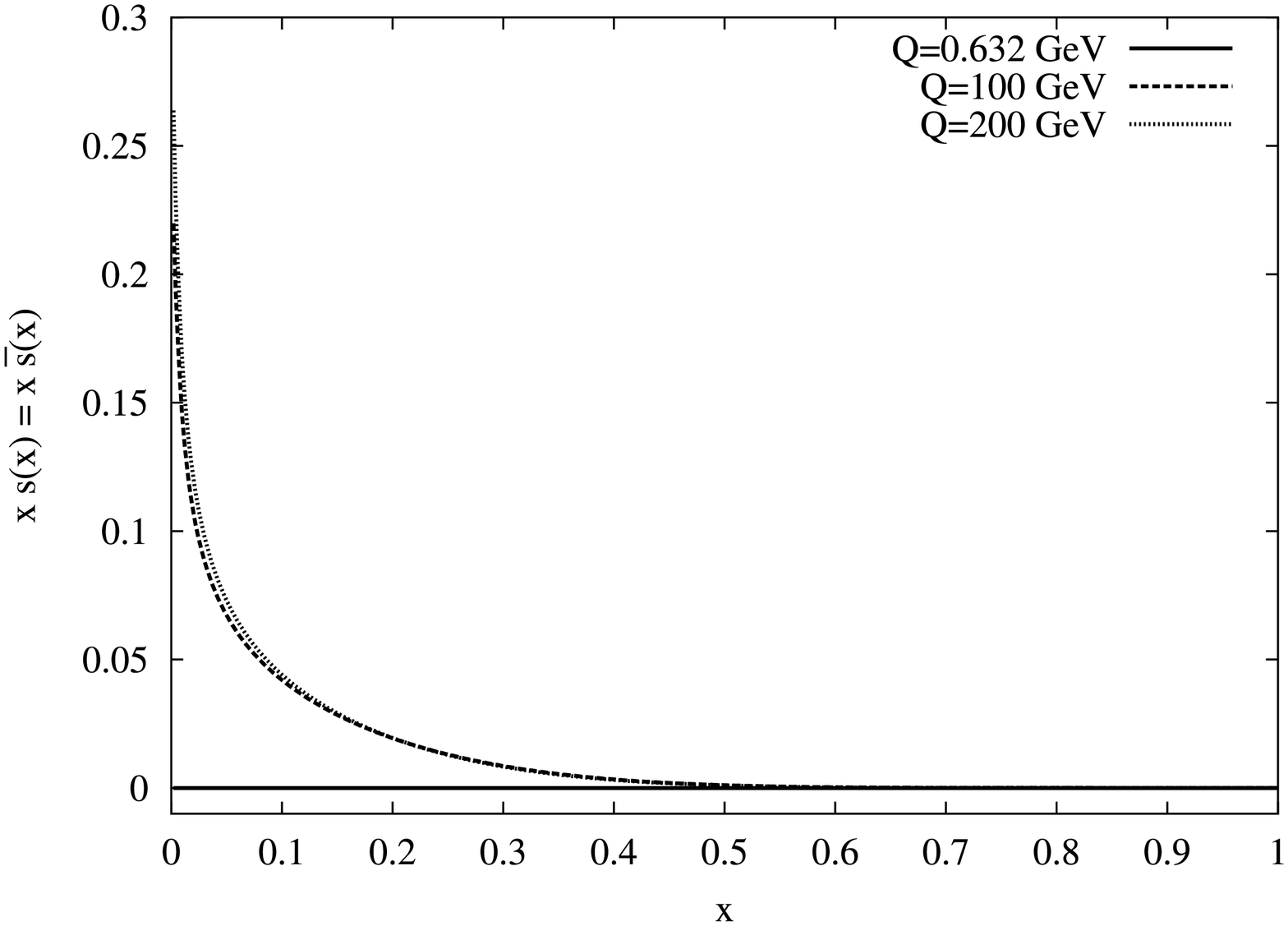}}} \par}

\caption{Evolution of \protect\( xs=x\overline{s}\protect \) versus \protect\( x\protect \)
at various \protect\( Q\protect \) values.}
\end{figure}

\begin{figure}[tbh]
{\centering \resizebox*{8cm}{!}{\rotatebox{-90}{\includegraphics{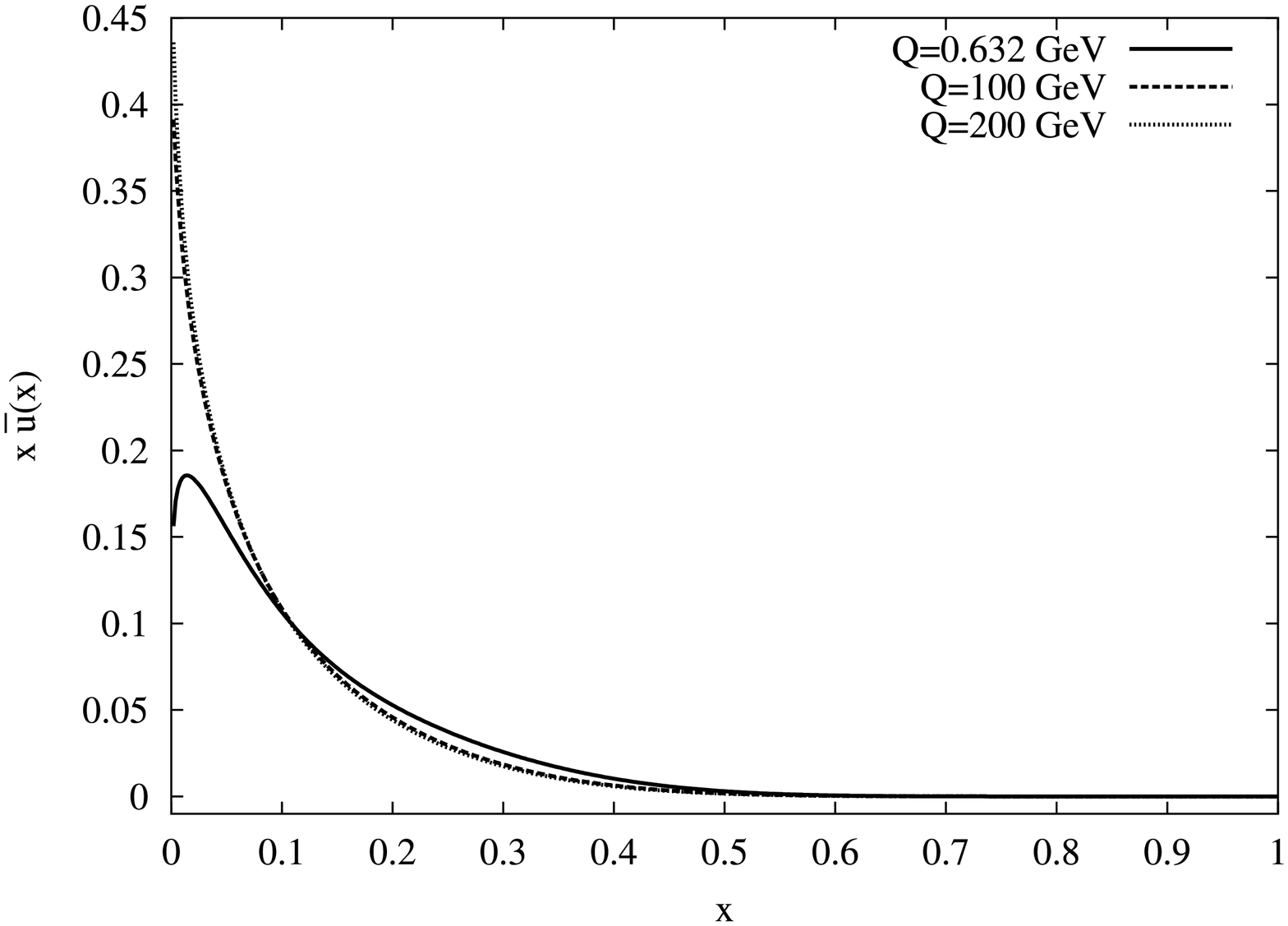}}} \par}

\caption{Evolution of the unpolarized antiquark up distribution \protect\( x\overline{u}\protect \)
versus \protect\( x\protect \) at various \protect\( Q\protect \)
values.}
\end{figure}

\begin{figure}[tbh]
{\centering \resizebox*{8cm}{!}{\rotatebox{-90}{\includegraphics{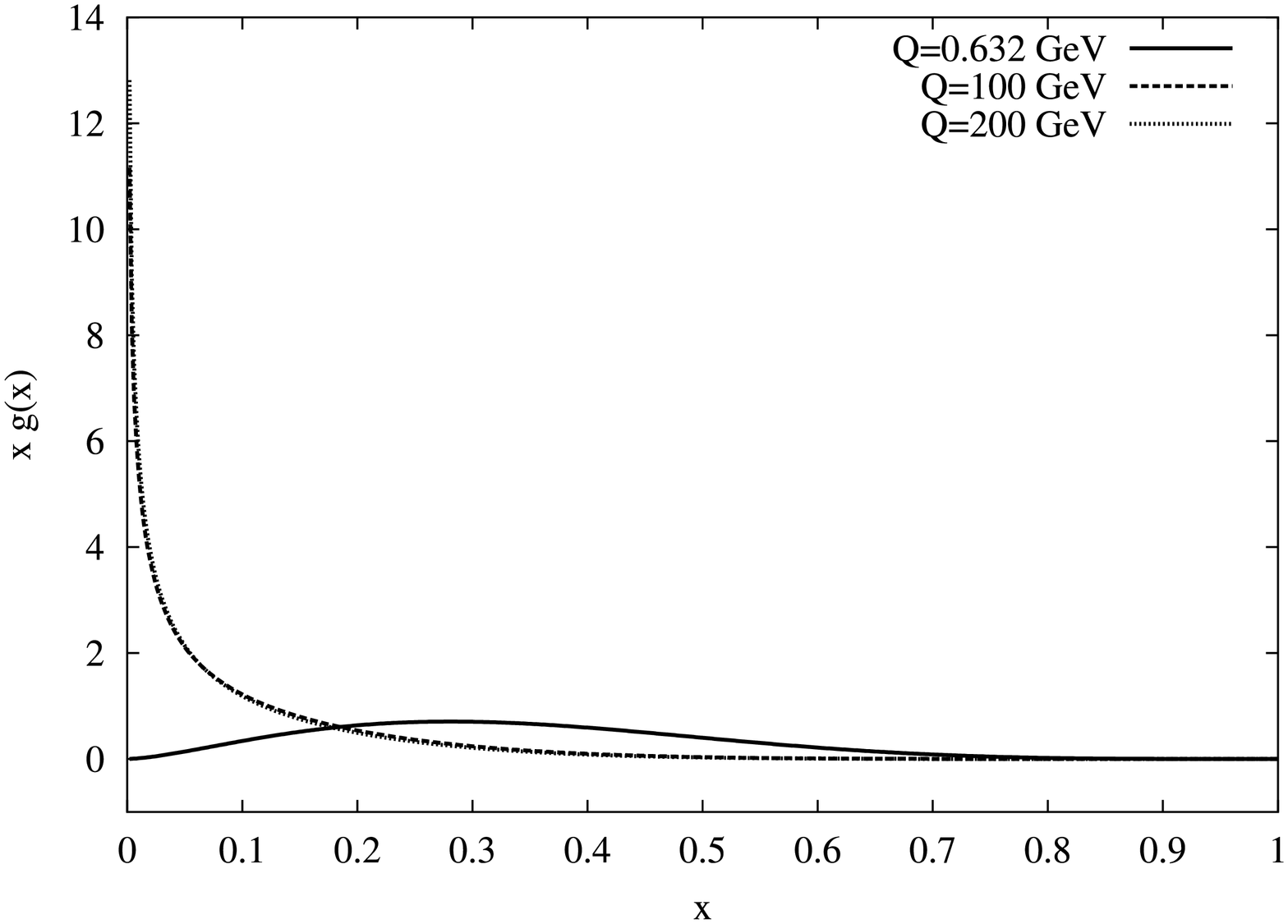}}} \par}

\caption{Evolution of the unpolarized gluon distribution \protect\( xg\protect \)
versus \protect\( x\protect \) at various \protect\( Q\protect \)
values.A}
\end{figure}

\begin{figure}[tbh]
{\centering \resizebox*{8cm}{!}{\rotatebox{-90}{\includegraphics{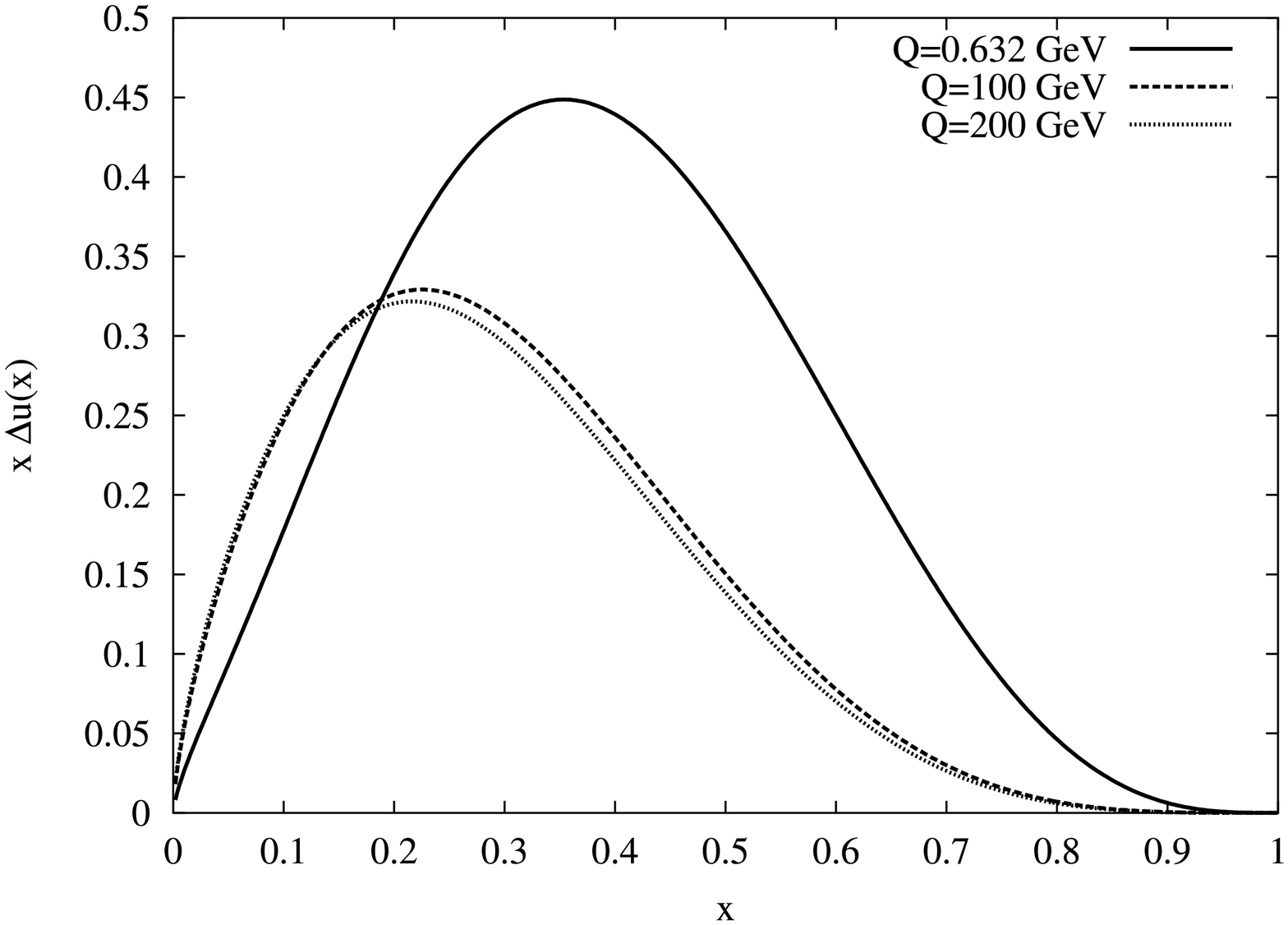}}} \par}

\caption{Evolution of the longitudinally polarized quark up distribution \protect\( x\Delta u\protect \)
versus \protect\( x\protect \) at various \protect\( Q\protect \)
values.\label{fig:Lu}}
\end{figure}

\begin{figure}[tbh]
{\centering \resizebox*{8cm}{!}{\rotatebox{-90}{\includegraphics{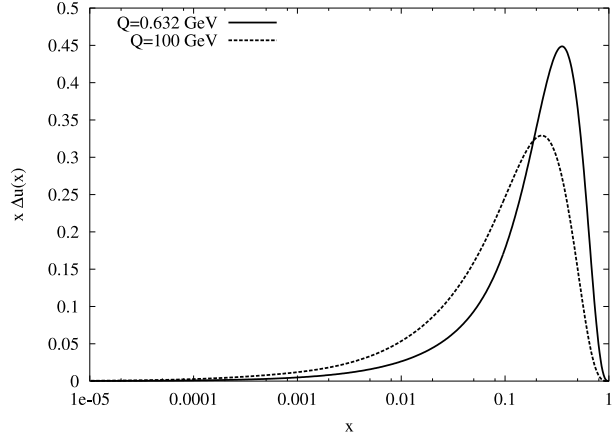}}} \par}

\caption{As in figure (\ref{fig:Lu}), but now the \protect\( x\protect \)-axis
is in logarithmic scale, to show the small-\protect\( x\protect \)
behavior.\label{fig:Lu_log}}
\end{figure}

\begin{figure}[tbh]
{\centering \resizebox*{8cm}{!}{\rotatebox{-90}{\includegraphics{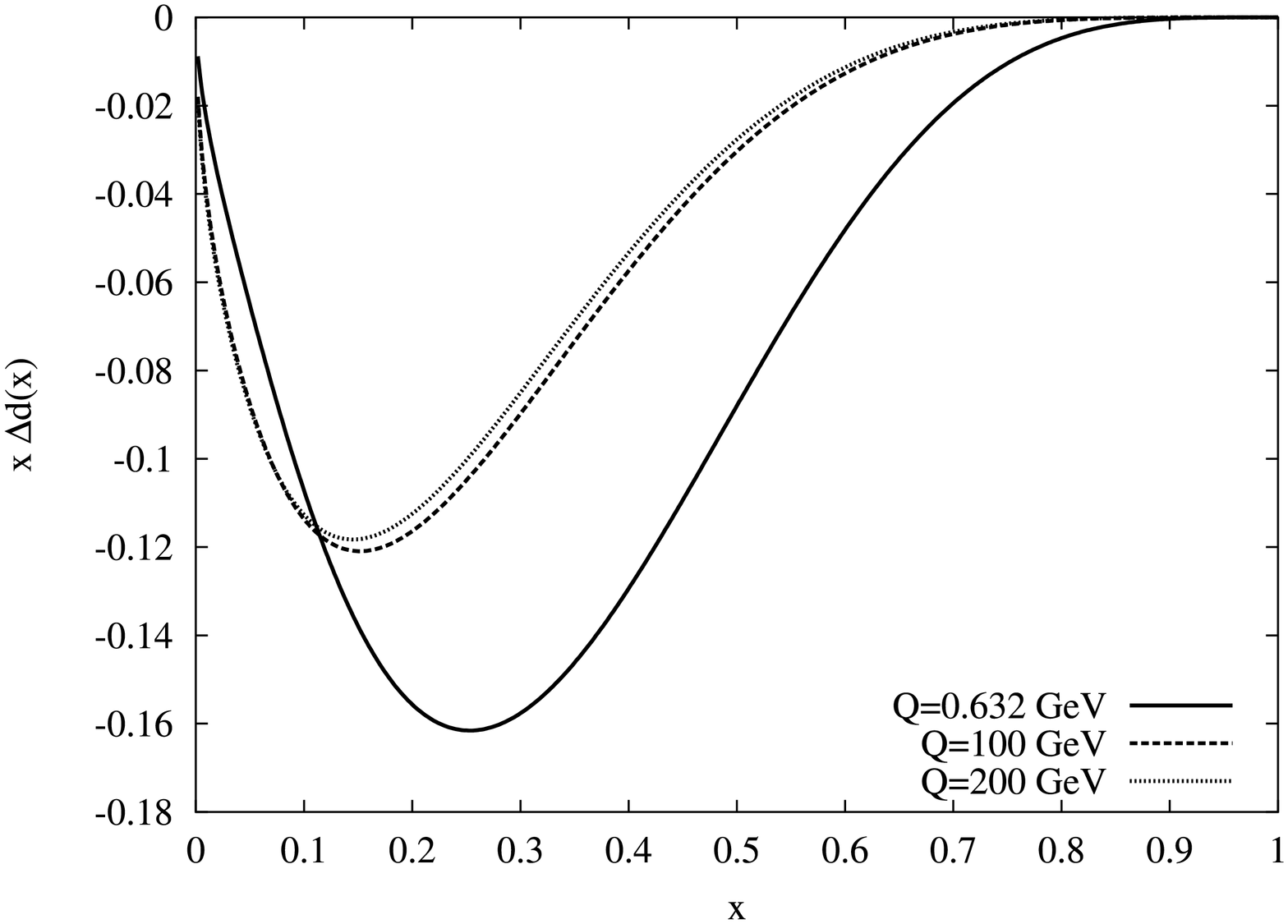}}} \par}

\caption{Evolution of \protect\( x\Delta d\protect \) versus \protect\( x\protect \)
at various \protect\( Q\protect \) values.}
\end{figure}

\begin{figure}[tbh]
{\centering \resizebox*{8cm}{!}{\rotatebox{-90}{\includegraphics{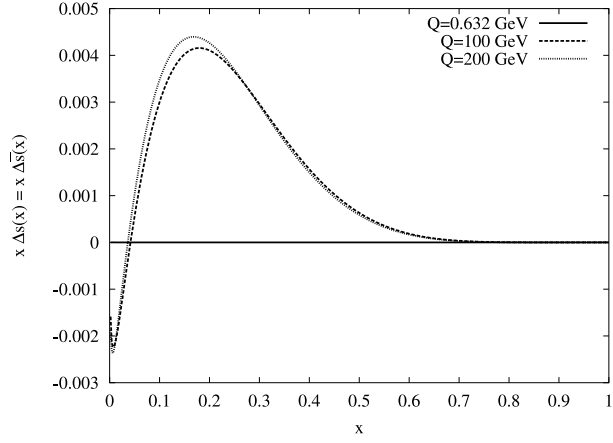}}} \par}

\caption{Evolution of \protect\( x\Delta s=x\Delta \overline{s}\protect \)
versus \protect\( x\protect \) at various \protect\( Q\protect \)
values.}
\end{figure}

\begin{figure}[tbh]
{\centering \resizebox*{8cm}{!}{\rotatebox{-90}{\includegraphics{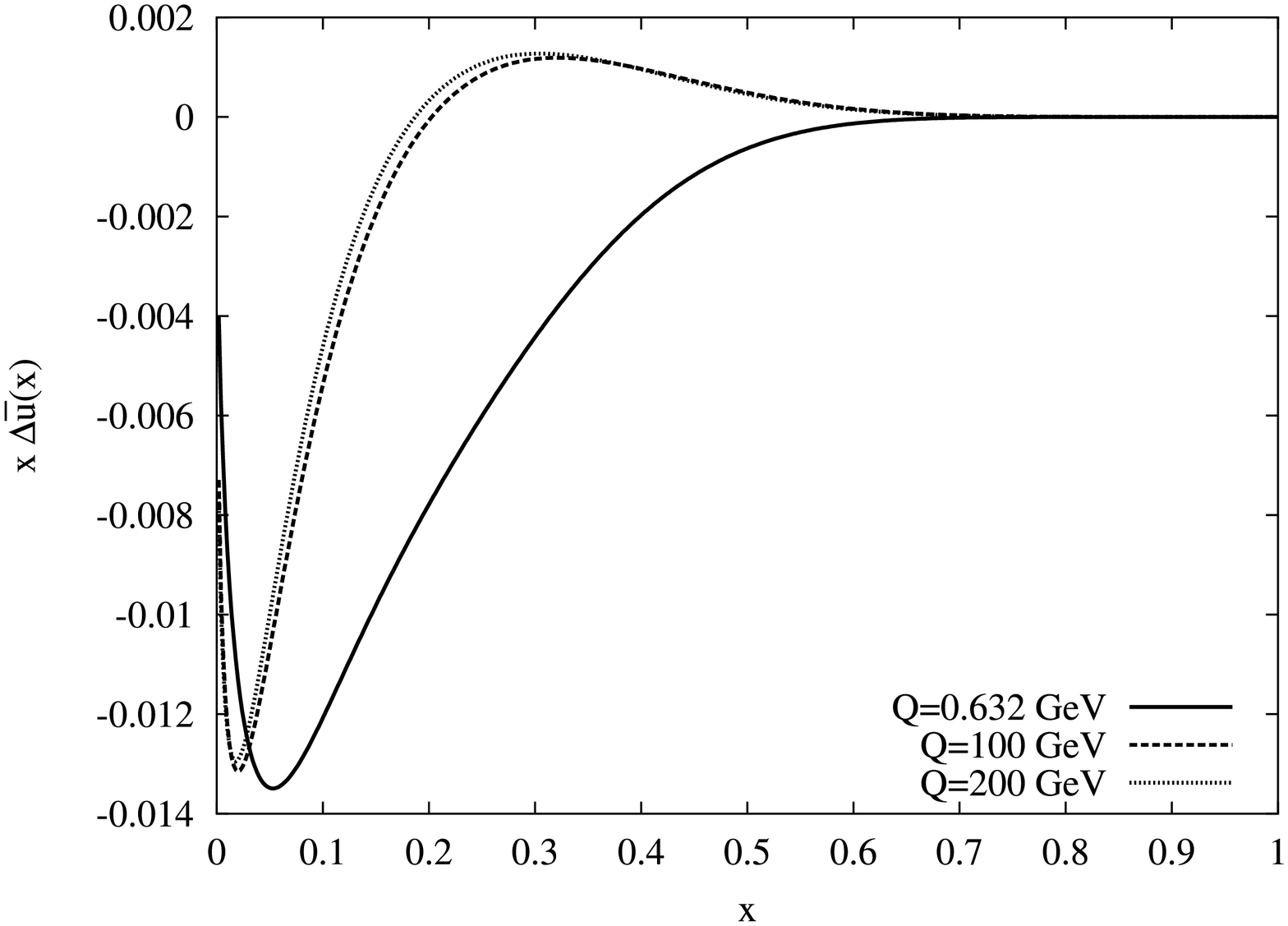}}} \par}

\caption{Evolution of the longitudinally polarized antiquark up distribution
\protect\( x\Delta \overline{u}\protect \) versus \protect\( x\protect \)
at various \protect\( Q\protect \) values.}
\end{figure}

\begin{figure}[tbh]
{\centering \resizebox*{8cm}{!}{\rotatebox{-90}{\includegraphics{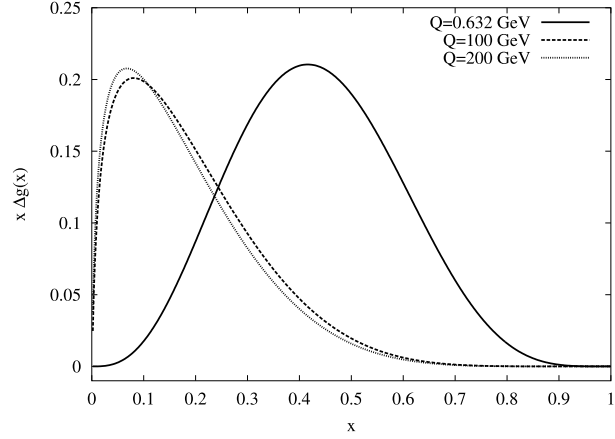}}} \par}

\caption{Evolution of the longitudinally polarized gluon distribution \protect\( x\Delta g\protect \)
versus \protect\( x\protect \) at various \protect\( Q\protect \)
values.\label{fig:Lg}}
\end{figure}

\begin{figure}[tbh]
{\centering \resizebox*{8cm}{!}{\rotatebox{-90}{\includegraphics{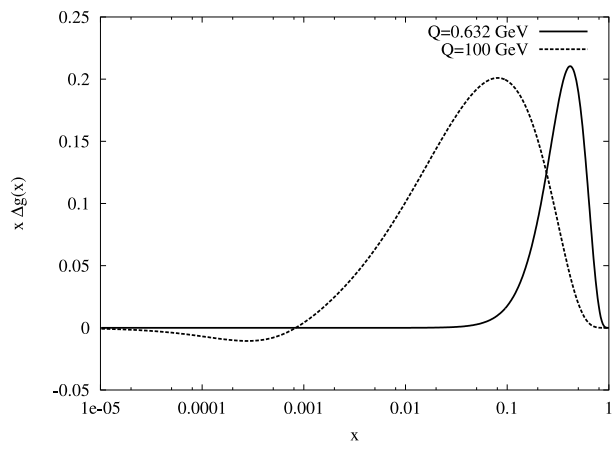}}} \par}

\caption{As in figure (\ref{fig:Lg}), but now the \protect\( x\protect \)-axis
is in logarithmic scale.\label{fig:Lg_log}}
\end{figure}

\begin{figure}[tbh]
{\centering \resizebox*{8cm}{!}{\rotatebox{-90}{\includegraphics{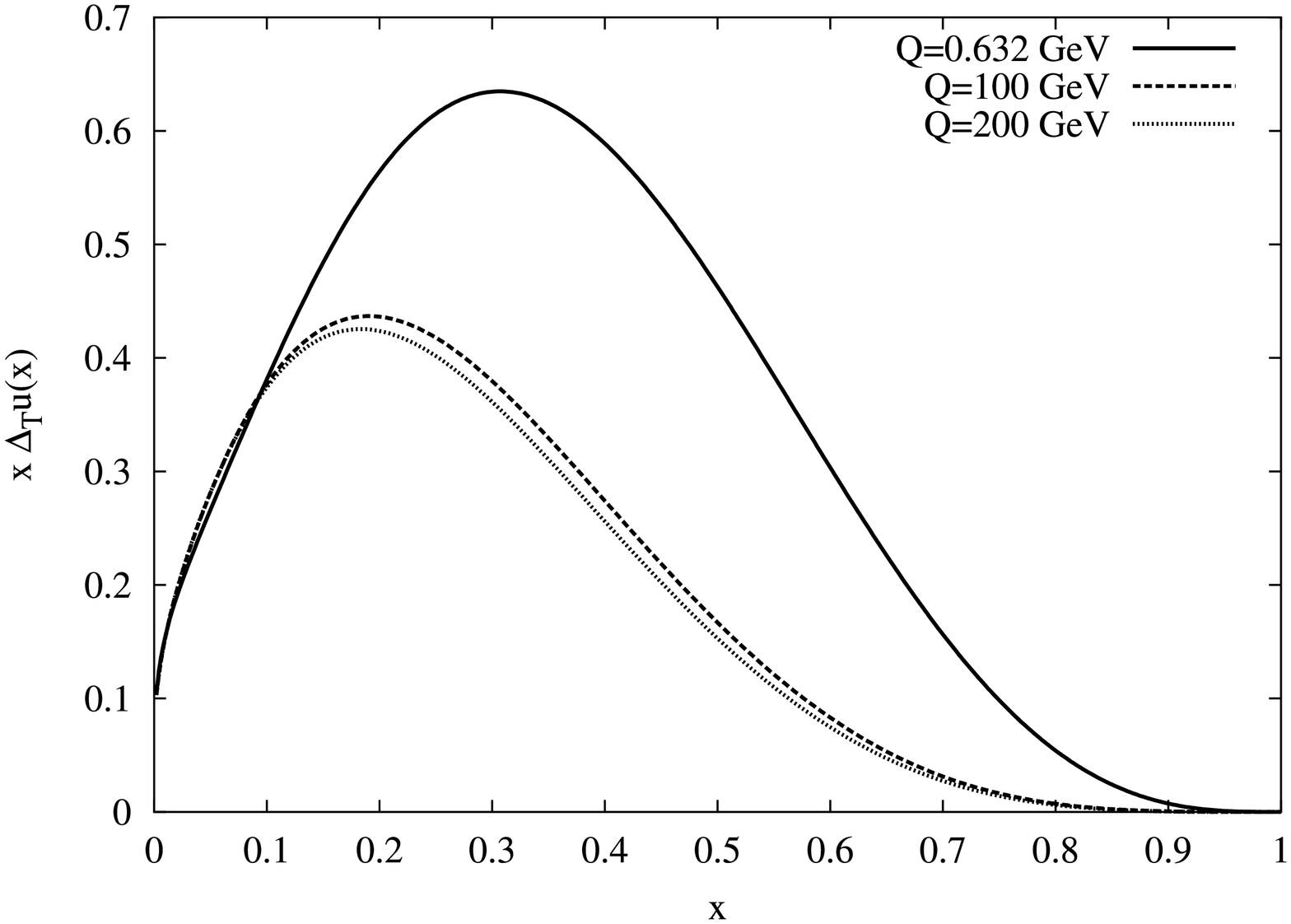}}} \par}

\caption{Evolution of the transversely polarized quark up distribution \protect\( x\Delta _{T}u\protect \)
versus \protect\( x\protect \) at various \protect\( Q\protect \)
values.}
\end{figure}

\begin{figure}[tbh]
{\centering \resizebox*{8cm}{!}{\rotatebox{-90}{\includegraphics{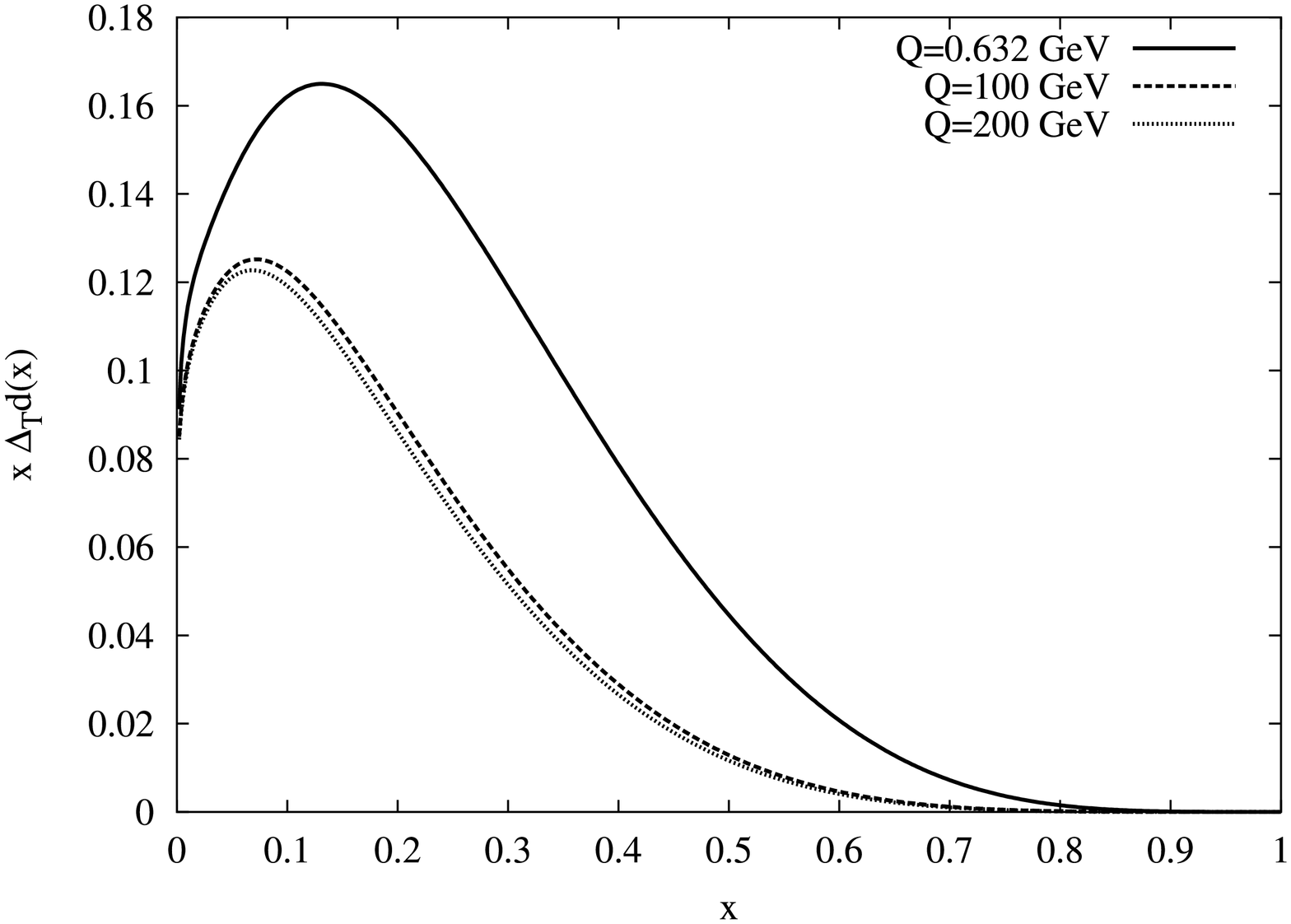}}} \par}

\caption{Evolution of \protect\( x\Delta _{T}d\protect \) versus \protect\( x\protect \)
at various \protect\( Q\protect \) values.}
\end{figure}

\begin{figure}[tbh]
{\centering \resizebox*{8cm}{!}{\rotatebox{-90}{\includegraphics{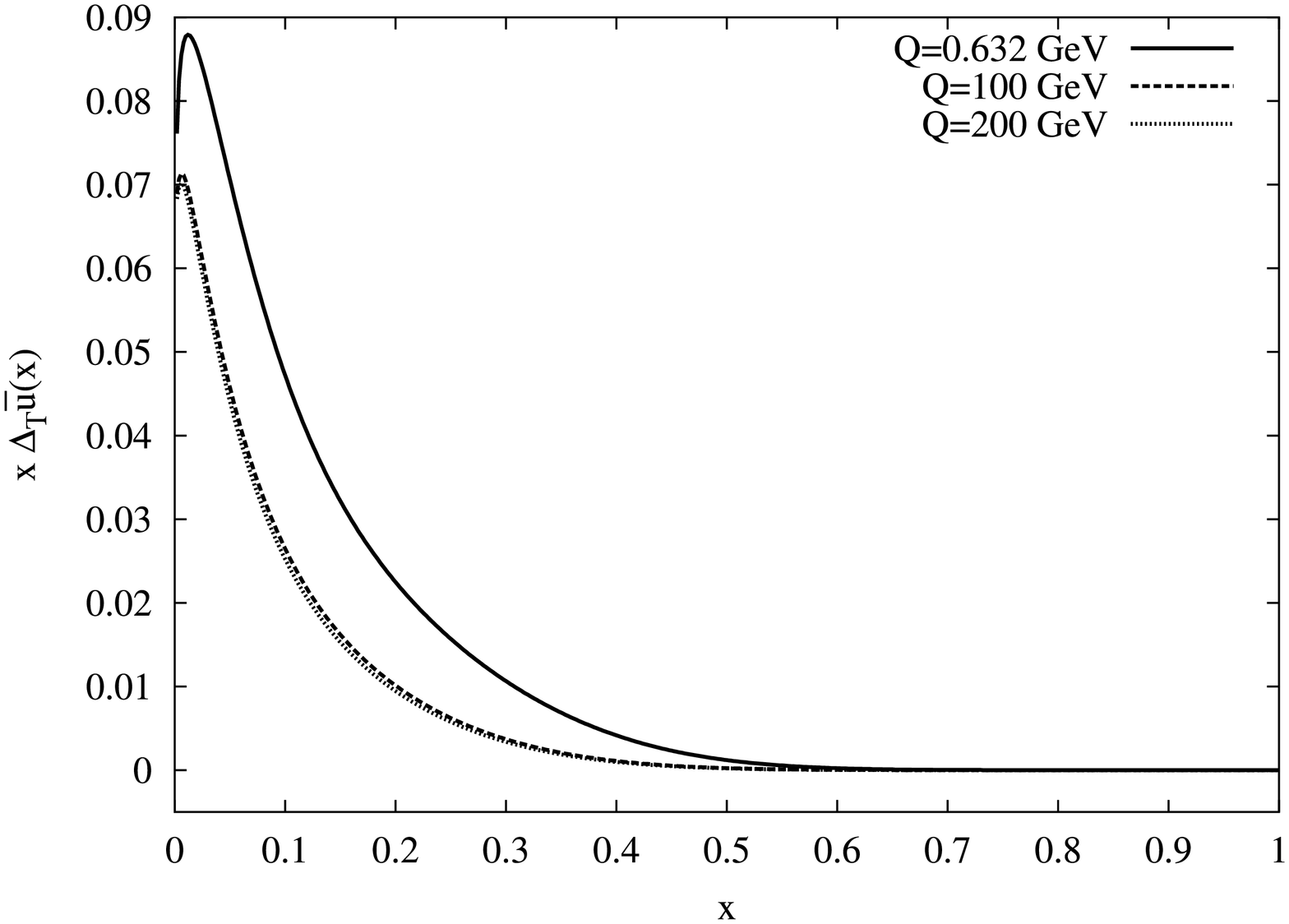}}} \par}

\caption{Evolution of the transversely polarized antiquark up distribution
\protect\( x\Delta _{T}\overline{u}\protect \) versus \protect\( x\protect \)
at various \protect\( Q\protect \) values.}
\end{figure}

\section{Conclusions}
We have illustrated and documented a method for solving 
in a rather fast way the NLO evolution equations 
for the parton distributions at leading twist. 
The advantages of the method compared to other implementations 
based on the inversion of the Mellin moments, 
as usually done in the case of QCD, 
are rather evident. 
We have also shown how Rossi's ansatz, originally formulated in the case 
of the photon structure function, relates to the solution of 
DGLAP equations formulated in terms of moments. 
The running time of the implementation is truly modest, even for a large 
number of iterations, and allows to get a very good accuracy.   
We hope to return with a discussion of the supersymmetric 
extension of the method and other applications in the near future.

\section*{Acknowledgments}
We thank the Phys. Dept. at the Univ. of Crete and especially 
Theodore Tomaras for hospitality and partial support under grant 
HPRN-CT-2000-00131 during this investigation 
and Marco Stratmann for correspondence. This work is partially supported 
by INFN (iniz. spec. BA21). 

\appendix

\section{Unpolarized kernels}

The reference for the unpolarized kernels is \cite{Petronzio}, rearranged for our 
purposes. 
We remind that the plus distribution is defined by\begin{equation}
\int _{0}^{1}\textrm{d}x\frac{f(x)}{(1-x)_{+}}=\int _{0}^{1}\textrm{d}x\frac{f(x)-f(1)}{1-x}
\end{equation}
and the Spence function is\begin{equation}
S_{2}(x)=-2\textrm{Li}_{2}(-x)-2\log x\log (1+x)+\frac{1}{2}\log ^{2}x-\frac{\pi ^{2}}{6},
\end{equation}
where the dilogarithm is defined by\begin{equation}
\textrm{Li}_{2}(x)=\int _{x}^{0}\frac{\log (1-t)}{t}\textrm{d}t.
\end{equation}

\begin{equation}
P^{(0)}_{NS^{-}}(x)=P^{(0)}_{NS^{+}}(x)=P^{(0)}_{qq}(x)=C_{F}\left[ \frac{2}{(1-x)_{+}}-1-x+\frac{3}{2}\delta (1-x)\right] 
\end{equation}
\begin{equation}
P^{(0)}_{qg}(x)=2T_{f}\left[ x^{2}+(1-x)^{2}\right] 
\end{equation}
\begin{equation}
P^{(0)}_{gq}(x)=C_{F}\left[ \frac{1+(1-x)^{2}}{x}\right] 
\end{equation}
\begin{equation}
P^{(0)}_{gg}(x)=2N_{C}\left[ \frac{1}{(1-x)_{+}}+\frac{1}{x}-2+x(1-x)\right] +\frac{\beta (0)}{2}\delta (1-x)
\end{equation}

\begin{eqnarray}
P^{(1)}_{NS^{-}}(x) & = & \left\{ \frac{C_{F}}{18}\left[ 162C_{F}(x-1)+4T_{f}(11x-1)+N_{C}(89-223x+3\pi ^{2}(1+x))\right] \right\} \nonumber \\
 &  & +\left\{ \frac{C_{F}\left[ 30C_{F}-23N_{C}+4T_{f}+12C_{F}x+(N_{C}-24C_{F}+4T_{f})x^{2}\right] }{6(x-1)}\right\} \log x\nonumber \\
 &  & +\left\{ \frac{C_{F}\left[ C_{F}-N_{C}-(C_{F}+N_{C})x^{2}\right] }{2(x-1)}\right\} \log ^{2}x\nonumber \\
 &  & +\left\{ \frac{2C_{F}^{2}(1+x^{2})}{x-1}\right\} \log x\log (1-x)\nonumber \\
 &  & -\left\{ \frac{C_{F}(2C_{F}-N_{C})(1+x^{2})}{1+x}\right\} S_{2}(x)\nonumber \\
 &  & -\left\{ \frac{C_{F}}{9}\left[ N_{C}(3\pi ^{2}-67)+20T_{f}\right] \right\} \frac{1}{(1-x)_{+}}\nonumber \\
 &  & +\left\{ \frac{C_{F}}{72}\left[ N_{C}(51+44\pi ^{2}-216\zeta (3))-4T_{f}(3+4\pi ^{2})\right. \right. \nonumber \\
 &  & \qquad \qquad \left. \left. +9C_{F}(3-4\pi ^{2}+48\zeta (3))\right] \right\} \delta (1-x)
\end{eqnarray}

\begin{eqnarray}
P^{(1)}_{NS^{+}}(x) & = & \left\{ \frac{C_{F}}{18}\left[ 18C_{F}(x-1)+4T_{f}(11x-1)+N_{C}(17-151x+3\pi ^{2}(1+x))\right] \right\} \nonumber \\
 &  & +\left\{ \frac{C_{F}\left[ 6C_{F}(1+2x)-(11N_{C}-4T_{f})(1+x^{2})\right] }{6(x-1)}\right\} \log x\nonumber \\
 &  & +\left\{ \frac{C_{F}\left[ C_{F}-N_{C}-(C_{F}+N_{C})x^{2}\right] }{2(x-1)}\right\} \log ^{2}x\nonumber \\
 &  & +\left\{ \frac{2C_{F}^{2}(1+x^{2})}{x-1}\right\} \log x\log (1-x)\nonumber \\
 &  & +\left\{ \frac{C_{F}(2C_{F}-N_{C})(1+x^{2})}{1+x}\right\} S_{2}(x)\nonumber \\
 &  & -\left\{ \frac{C_{F}}{9}\left[ N_{C}(3\pi ^{2}-67)+20T_{f}\right] \right\} \frac{1}{(1-x)_{+}}\nonumber \\
 &  & +\left\{ \frac{C_{F}}{72}\left[ N_{C}(51+44\pi ^{2}-216\zeta (3))-4T_{f}(3+4\pi ^{2})\right. \right. \nonumber \\
 &  & \qquad \qquad \left. \left. +9C_{F}(3-4\pi ^{2}+48\zeta (3))\right] \right\} \delta (1-x)
\end{eqnarray}
\begin{eqnarray}
P^{(1)}_{qq}(x) & = & \frac{C_{F}}{18x}\left\{ x\left[ 18C_{F}(x-1)+N_{C}\left( 17-151x+3\pi ^{2}(1+x)\right) \right] \right. \nonumber \\
 &  & \qquad \quad \left. +4T_{f}\left[ 20-x\left( 19+x(56x-65)\right) \right] \right\} \nonumber \\
 &  & +\left\{ \frac{C_{F}\left[ 6C_{F}(1+2x)-11N_{C}(1+x^{2})+8T_{f}\left( 2x(2x(1+x)-3)-1\right) \right] }{6(x-1)}\right\} \log x\nonumber \\
 &  & +\left\{ \frac{C_{F}\left[ C_{F}-N_{C}+4T_{f}-(C_{F}+N_{C}+4T_{f})x^{2}\right] }{2(x-1)}\right\} \log ^{2}x\nonumber \\
 &  & +\left\{ \frac{2C_{F}^{2}(1+x^{2})}{x-1}\right\} \log x\log (1-x)\nonumber \\
 &  & +\left\{ \frac{C_{F}(2C_{F}-N_{C})(1+x^{2})}{1+x}\right\} S_{2}(x)\nonumber \\
 &  & -\left\{ \frac{C_{F}}{9}\left[ N_{C}(3\pi ^{2}-67)+20T_{f}\right] \right\} \frac{1}{(1-x)_{+}}\nonumber \\
 &  & +\left\{ \frac{C_{F}}{72}\left[ N_{C}(51+44\pi ^{2}-216\zeta (3))-4T_{f}(3+4\pi ^{2})\right. \right. \nonumber \\
 &  & \qquad \qquad \left. \left. +9C_{F}(3-4\pi ^{2}+48\zeta (3))\right] \right\} \delta (1-x)
\end{eqnarray}
\begin{eqnarray}
P^{(1)}_{qg}(x) & = & \left\{ \frac{1}{9x}\left[ T_{f}\left( 3C_{F}x(42-87x+60x^{2}-\pi ^{2}(2+4(x-1)x))\right. \right. \right. \nonumber \\
 &  & \qquad \qquad \left. \left. \left. +N_{C}(40+x(450x-36-436x^{2}+\pi ^{2}(3+6(x-1)x)))\right) \right] \right\} \nonumber \\
 &  & +\left\{ \frac{T_{f}}{3}\left[ 6N_{C}+8N_{C}x(6+11x)+3C_{F}(3-4x+8x^{2})\right] \right\} \log x\nonumber \\
 &  & +\left\{ 8(C_{F}-N_{C})T_{f}(1-x)x\right\} \log (1-x)\nonumber \\
 &  & +\left\{ T_{f}\left[ C_{F}(1-2x+4x^{2})-N_{C}(3+2x(3+x))\right] \right\} \log ^{2}x\nonumber \\
 &  & +\left\{ 2(C_{F}-N_{C})T_{f}\left[ 1+2(x-1)x\right] \right\} \log ^{2}(1-x)\nonumber \\
 &  & -\left\{ 4C_{F}T_{f}\left[ 1+2(x-1)x\right] \right\} \log x\log (1-x)\nonumber \\
 &  & +\left\{ 2N_{C}T_{f}\left[ 1+2x(1+x)\right] \right\} S_{2}(x)
\end{eqnarray}
\begin{eqnarray}
P^{(1)}_{gq}(x) & = & \left\{ \frac{1}{18x}\left[ C_{F}\left( N_{C}(18-3\pi ^{2}(2+(x-2)x)+2x(19+x(37+44x)))\right. \right. \right. \nonumber \\
 &  & \qquad \qquad \left. \left. \left. -9C_{F}x(5+7x)-16T_{f}(5+x(4x-5))\right) \right] \right\} \nonumber \\
 &  & +\left\{ \frac{C_{F}}{6}\left[ 3C_{F}(4+7x)-2N_{C}\left( 36+x(15+8x)\right) \right] \right\} \log x\nonumber \\
 &  & +\left\{ \frac{C_{F}}{3x}\left[ N_{C}\left( 22+x(17x-22)\right) -4T_{f}\left( 2+(x-2)x\right) \right. \right. \nonumber \\
 &  & \qquad \qquad \left. \left. -3C_{F}\left( 6+x(5x-6)\right) \right] \right\} \log (1-x)\nonumber \\
 &  & +\left\{ \frac{C_{F}}{2x}\left[ C_{F}(x-2)x+N_{C}\left( 2+3x(2+x)\right) \right] \right\} \log ^{2}x\nonumber \\
 &  & +\left\{ \frac{C_{F}(N_{C}-C_{F})\left[ 2+(x-2)x\right] }{x}\right\} \log ^{2}(1-x)\nonumber \\
 &  & -\left\{ \frac{2C_{F}N_{C}\left( 2+(x-2)x\right) }{x}\right\} \log x\log (1-x)\nonumber \\
 &  & -\left\{ \frac{C_{F}N_{C}\left( 2+x(2+x)\right) }{x}\right\} S_{2}(x)
\end{eqnarray}
\begin{eqnarray}
P^{(1)}_{gg}(x) & = & \left\{ \frac{1}{18x}\left[ 24C_{F}T_{f}(x-1)\left( x(11+5x)-1\right) +4N_{C}T_{f}\left( x(29+x(23x-19))-23\right) \right. \right. \nonumber \\
 &  & \qquad \quad \left. \left. +N_{C}^{2}\left( 6\pi ^{2}(x(2+(x-1)x)-1)-x(25+109x)\right) \right] \right\} \nonumber \\
 &  & +\left\{ \frac{N_{C}^{2}\left[ 11(1-4x)x-25\right] -4N_{C}T_{f}(1+x)-6C_{F}T_{f}(3+5x)}{3}\right\} \log x\nonumber \\
 &  & +\left\{ \frac{2C_{F}T_{f}x(x^{2}-1)+N_{C}^{2}\left[ 1+x\left( 2+x(3+(x-6)x)\right) \right] }{(1-x)x}\right\} \log ^{2}x\nonumber \\
 &  & +\left\{ \frac{4N_{C}^{2}\left[ 1+(x-1)x\right] ^{2}}{(x-1)x}\right\} \log x\log (1-x)\nonumber \\
 &  & -\left\{ \frac{2N_{C}^{2}\left( 1+x+x^{2}\right) ^{2}}{x(1+x)}\right\} S_{2}(x)\nonumber \\
 &  & -\left\{ \frac{N_{C}}{9}\left[ N_{C}(3\pi ^{2}-67)+20T_{f}\right] \right\} \frac{1}{(1-x)_{+}}\nonumber \\
 &  & +\left\{ \frac{N_{C}}{3}\left[ N_{C}(8+9\zeta (3))-4T_{f}\right] -C_{F}T_{f}\right\} \delta (1-x)
\end{eqnarray}

\section{Longitudinally polarized kernels}

The reference for the longitudinally polarized kernels is \cite{Vogelsang96}.

\begin{equation}
\Delta P^{(0)}_{NS^{-}}(x)=\Delta P^{(0)}_{NS^{+}}(x)=\Delta P^{(0)}_{qq}(x)=C_{F}\left[ \frac{2}{(1-x)_{+}}-1-x+\frac{3}{2}\delta (1-x)\right] 
\end{equation}
\begin{equation}
\Delta P_{qg}^{(0)}(x)=2T_{f}(2x-1)
\end{equation}
\begin{equation}
\Delta P_{gq}^{(0)}(x)=C_{F}(2-x)
\end{equation}
\begin{equation}
\Delta P_{gg}^{(0)}(x)=2N_{C}\left[ \frac{1}{(1-x)_{+}}-2x+1\right] +\frac{\beta (0)}{2}\delta (1-x)
\end{equation}

\begin{equation}
\Delta P^{(1)}_{NS^{-}}(x)=P^{(1)}_{NS^{+}}(x)
\end{equation}
\begin{equation}
\Delta P^{(1)}_{NS^{+}}(x)=P^{(1)}_{NS^{-}}(x)
\end{equation}
\begin{eqnarray}
\Delta P^{(1)}_{qq}(x) & = & \left\{ \frac{C_{F}}{18}\left[ 162C_{F}(x-1)+8T_{f}(4+z)+N_{C}\left( 89-223x+3\pi ^{2}(1+x)\right) \right] \right\} \nonumber \\
 &  & +\left\{ \frac{C_{F}}{6(x-1)}\left[ N_{C}(x^{2}-23)-6C_{F}(4x^{2}-2x-5)\right. \right. \nonumber \\
 &  & \qquad \qquad \qquad \left. \left. +8T_{f}\left( 2+x(5x-6)\right) \right] \right\} \log x\nonumber \\
 &  & +\left\{ \frac{C_{F}\left[ C_{F}-N_{C}+4T_{f}-(C_{F}+N_{C}+4T_{f})x^{2}\right] }{2(x-1)}\right\} \log ^{2}x\nonumber \\
 &  & +\left\{ \frac{2C_{F}^{2}(1+x^{2})}{x-1}\right\} \log x\log (1-x)\nonumber \\
 &  & -\left\{ \frac{C_{F}(2C_{F}-N_{C})(1+x^{2})}{1+x}\right\} S_{2}(x)\nonumber \\
 &  & -\left\{ \frac{C_{F}}{9}\left[ N_{C}(3\pi ^{2}-67)+20T_{f}\right] \right\} \frac{1}{(1-x)_{+}}\nonumber \\
 &  & +\left\{ \frac{C_{F}}{72}\left[ N_{C}(51+44\pi ^{2}-216\zeta (3))-4T_{f}(3+4\pi ^{2})\right. \right. \nonumber \\
 &  & \qquad \qquad \left. \left. +9C_{F}(3-4\pi ^{2}+48\zeta (3))\right] \right\} \delta (1-x)
\end{eqnarray}
\begin{eqnarray}
\Delta P^{(1)}_{qg}(x) & = & \left\{ \frac{T_{f}}{3}\left[ C_{F}\left( \pi ^{2}(2-4x)-66+81x\right) +N_{C}\left( 72-66x+\pi ^{2}(2x-1)\right) \right] \right\} \nonumber \\
 &  & +\left\{ T_{f}\left[ 2N_{C}(1+8x)-9C_{F}\right] \right\} \log x\nonumber \\
 &  & +\left\{ 8(N_{C}-C_{F})T_{f}(x-1)\right\} \log (1-x)\nonumber \\
 &  & +\left\{ T_{f}\left[ C_{F}(2x-1)-3N_{C}(1+2x)\right] \right\} \log ^{2}x\nonumber \\
 &  & +\left\{ 2(C_{F}-N_{C})T_{f}(2x-1)\right\} \log ^{2}(1-x)\nonumber \\
 &  & +\left\{ 4C_{F}T_{f}(1-2x)\right\} \log x\log (1-x)\nonumber \\
 &  & +\left\{ 2N_{C}T_{f}(1+2x)\right\} S_{2}(x)
\end{eqnarray}
\begin{eqnarray}
\Delta P^{(1)}_{gq}(x) & = & \left\{ \frac{C_{F}}{18}\left[ 9C_{F}(8x-17)-8T_{f}(4+x)+N_{C}\left( 82+3\pi ^{2}(x-2)+70x\right) \right] \right\} \nonumber \\
 &  & +\left\{ \frac{C_{F}}{2}\left[ N_{C}(8-26x)+C_{F}(x-4)\right] \right\} \log x\nonumber \\
 &  & +\left\{ \frac{C_{F}}{3}\left[ 4T_{f}(x-2)-3C_{F}(2+x)+N_{C}(10+x)\right] \right\} \log (1-x)\nonumber \\
 &  & +\left\{ \frac{C_{F}}{2}\left[ 3N_{C}(2+x)-C_{F}(x-2)\right] \right\} \log ^{2}x\nonumber \\
 &  & +\left\{ C_{F}(C_{F}-N_{C})(x-2)\right\} \log ^{2}(1-x)\nonumber \\
 &  & +\left\{ 2C_{F}N_{C}(x-2)\right\} \log x\log (1-x)\nonumber \\
 &  & -\left\{ C_{F}N_{C}(2+x)\right\} S_{2}(x)
\end{eqnarray}
\begin{eqnarray}
\Delta P^{(1)}_{gg}(x) & = & \left\{ \frac{1}{18}\left[ 180C_{F}T_{f}(x-1)+8N_{C}T_{f}(19x-4)+N_{C}^{2}\left( 6\pi ^{2}(1+2x)-305-97x\right) \right] \right\} \nonumber \\
 &  & +\left\{ \frac{1}{3}\left[ N_{C}^{2}(29-67x)+6C_{F}T_{f}(x-5)-4N_{C}T_{f}(1+x)\right] \right\} \log x\nonumber \\
 &  & +\left\{ \frac{N_{C}^{2}(2x^{2}+x-4)-2C_{F}T_{f}(x^{2}-1)}{x-1}\right\} \log ^{2}x\nonumber \\
 &  & +\left\{ \frac{4N_{C}^{2}x(2x-1)}{x-1}\right\} \log x\log (1-x)\nonumber \\
 &  & -\left\{ \frac{2N_{C}^{2}x(1+2x)}{1+x}\right\} S_{2}(x)\nonumber \\
 &  & -\left\{ \frac{N_{C}}{9}\left[ N_{C}(3\pi ^{2}-67)+20T_{f}\right] \right\} \frac{1}{(1-x)_{+}}\nonumber \\
 &  & +\left\{ \frac{1}{3}\left[ N_{C}^{2}(8+9\zeta (3))-3C_{F}T_{f}-4N_{C}T_{f}\right] \right\} \delta (1-x)
\end{eqnarray}

\section{Kernels in the helicity basis}
The expression of the kernels in the helicity basis given below are 
obtained combining the NLO computations of \cite{Petronzio,Vogelsang96}

\begin{eqnarray}
P^{(1)}_{NS^{-},++}(x) & = & \left\{ \frac{C_{F}}{18}\left[ 90C_{F}(x-1)+4T_{f}(11x-1)+N_{C}(53-187x+3\pi ^{2}(1+x))\right] \right\} \nonumber \\
 &  & +\left\{ \frac{C_{F}\left[ 6C_{F}(3-2(x-1)x)+4T_{f}(1+x^{2})-N_{C}(17+5x^{2})\right] }{6(x-1)}\right\} \log x\nonumber \\
 &  & +\left\{ \frac{C_{F}\left[ C_{F}-N_{C}-(C_{F}+N_{C})x^{2}\right] }{2(x-1)}\right\} \log ^{2}x\nonumber \\
 &  & +\left\{ \frac{2C_{F}^{2}(1+x^{2})}{x-1}\right\} \log x\log (1-x)\nonumber \\
 &  & +\left\{ -\frac{C_{F}}{9}\left[ N_{C}(3\pi ^{2}-67)+20T_{f}\right] \right\} \frac{1}{(1-x)_{+}}\nonumber \\
 &  & +\left\{ C_{F}\left[ \frac{N_{C}(51+44\pi ^{2})-4T_{f}(3+4\pi ^{2})}{72}-3N_{C}\zeta (3)\right. \right. \nonumber \\
 &  & \left. \left. \qquad +C_{F}\left( \frac{3}{8}-\frac{\pi ^{2}}{2}+6\zeta (3)\right) \right] \right\} \delta (1-x)
\end{eqnarray}
\begin{eqnarray}
P^{(1)}_{NS^{-},+-}(x) & = & \left\{ 2C_{F}(2C_{F}-N_{C})(x-1)\right\} \nonumber \\
 &  & +\left\{ C_{F}(N_{C}-2C_{F})(1+x)\right\} \log x\nonumber \\
 &  & +\left\{ \frac{C_{F}(N_{C}-2C_{F})(1+x^{2})}{1+x}\right\} S_{2}(x)
\end{eqnarray}
\begin{equation}
P^{(1)}_{NS^{+},++}(x)=P^{(1)}_{NS^{-},++}(x)
\end{equation}
\begin{equation}
P^{(1)}_{NS^{+},+-}(x)=-P^{(1)}_{NS^{-},+-}(x)
\end{equation}
\begin{eqnarray}
P^{(1)}_{qq,++}(x) & = & \left\{ \frac{C_{F}}{18x}\left[ 2T_{f}(20-(x-1)x(56x-11))\right. \right. \nonumber \\
 &  & \left. \left. \qquad \qquad +x(90C_{F}(x-1)+N_{C}(53-187x+3\pi ^{2}(1+x)))\right] \right\} \nonumber \\
 &  & +\left\{ \frac{C_{F}}{6(x-1)}\left[ 6C_{F}(3-2(x-1)x)-N_{C}(17+5x^{2})\right. \right. \nonumber \\
 &  & \left. \left. \qquad \qquad \qquad +4T_{f}(1+x(x(9+4x)-12))\right] \right\} \log x\nonumber \\
 &  & +\left\{ \frac{C_{F}\left[ C_{F}-N_{C}+4T_{f}-(C_{F}+N_{C}+4T_{F})x^{2}\right] }{2(x-1)}\right\} \log ^{2}x\nonumber \\
 &  & +\left\{ \frac{2C_{F}^{2}(1+x^{2})}{x-1}\right\} \log x\log (1-x)\nonumber \\
 &  & +\left\{ -\frac{C_{F}}{9}\left[ N_{C}(3\pi ^{2}-67)+20T_{f}\right] \right\} \frac{1}{(1-x)_{+}}\nonumber \\
 &  & +\left\{ \frac{C_{F}}{72}\left[ N_{C}(51+44\pi ^{2}-216\zeta (3))-4T_{f}(3+4\pi ^{2})\right. \right. \nonumber \\
 &  & \left. \left. \qquad \qquad +9C_{F}(3-4\pi ^{2}+48\zeta (3))\right] \right\} \delta (1-x)
\end{eqnarray}
\begin{eqnarray}
P^{(1)}_{qq,+-}(x) & = & \left\{ \frac{C_{F}(1-x)}{9x}\left[ 18(2C_{F}-N_{C})x+T_{f}(20-7x+56x^{2})\right] \right\} \nonumber \\
 &  & +\left\{ \frac{C_{F}}{3}\left[ 6C_{F}(1+x)-3N_{C}(1+x)+2T_{f}(3+x(3+4x))\right] \right\} \log x\nonumber \\
 &  & +\left\{ \frac{C_{F}(2C_{F}-N_{C})(1+x^{2})}{1+x}\right\} S_{2}(x)
\end{eqnarray}
\begin{eqnarray}
P^{(1)}_{qg,++}(x) & = & \left\{ \frac{T_{f}}{9x}\left[ N_{C}(20+x(90+x(126+(3\pi ^{2}-218)x)))\right. \right. \nonumber \\
 &  & \left. \left. \qquad -3C_{F}x(12-x(2(15-\pi ^{2})x-3))\right] \right\} \nonumber \\
 &  & +\left\{ \frac{T_{f}}{3}\left[ 6N_{C}+4N_{C}x(12+11x)-3C_{F}(3+2x-4x^{2})\right] \right\} \log x\nonumber \\
 &  & +\left\{ 4T_{f}(C_{F}-N_{C})(1-x^{2})\right\} \log (1-x)\nonumber \\
 &  & +\left\{ T_{f}\left[ 2C_{F}x^{2}-N_{C}(3+x(6+x))\right] \right\} \log ^{2}x\nonumber \\
 &  & +\left\{ 2T_{f}(C_{F}-N_{C})x^{2}\right\} \log ^{2}(1-x)\nonumber \\
 &  & +\left\{ -4C_{F}T_{f}x^{2}\right\} \log x\log (1-x)\nonumber \\
 &  & +\left\{ 2N_{C}T_{f}(1+x)^{2}\right\} S_{2}(x)
\end{eqnarray}
\begin{eqnarray}
P^{(1)}_{qg,+-}(x) & = & \left\{ \frac{T_{f}(x-1)}{9x}\left[ 6C_{F}x(15x-27-\pi ^{2}(x-1))\right. \right. \nonumber \\
 &  & \left. \left. \qquad \qquad \qquad -N_{C}(20-(106+3\pi ^{2}(x-1)-218x)x)\right] \right\} \nonumber \\
 &  & +\left\{ \frac{2}{3}T_{f}\left[ 22N_{C}x^{2}+3C_{F}(3+x(2x-1))\right] \right\} \log x\nonumber \\
 &  & +\left\{ 4T_{f}(N_{C}-C_{F})(1-x)^{2}\right\} \log (1-x)\nonumber \\
 &  & +\left\{ T_{f}\left[ C_{F}(1+2(x-1)x)-N_{C}x^{2}\right] \right\} \log ^{2}x\nonumber \\
 &  & +\left\{ 2T_{f}(C_{F}-N_{C})(1-x)^{2}\right\} \log ^{2}(1-x)\nonumber \\
 &  & +\left\{ -4C_{F}T_{f}(1-x)^{2}\right\} \log x\log (1-x)\nonumber \\
 &  & +\left\{ 2N_{C}T_{f}x^{2}\right\} S_{2}(x)
\end{eqnarray}
\begin{eqnarray}
P^{(1)}_{gq,++}(x) & = & \left\{ \frac{C_{F}}{36x}\left[ 9C_{F}(x-22)x-8T_{f}(10+3x(3x-2))\right. \right. \nonumber \\
 &  & \left. \left. \qquad \qquad +2N_{C}(9-3\pi ^{2}+4x(15+x(18+11x)))\right] \right\} \nonumber \\
 &  & +\left\{ \frac{C_{F}}{3}\left[ 6C_{F}x-N_{C}(12+x(27+4x))\right] \right\} \log x\nonumber \\
 &  & +\left\{ \frac{C_{F}}{3x}\left[ N_{C}(11-3(2-3x)x)-4T_{f}-3C_{F}(3+x(3x-2))\right] \right\} \log (1-x)\nonumber \\
 &  & +\left\{ \frac{C_{F}N_{C}(1+3x(2+x))}{2x}\right\} \log ^{2}x\nonumber \\
 &  & +\left\{ \frac{C_{F}(N_{C}-C_{F})}{x}\right\} \log ^{2}(1-x)\nonumber \\
 &  & +\left\{ -\frac{2C_{F}N_{C}}{x}\right\} \log x\log (1-x)\nonumber \\
 &  & +\left\{ -\frac{C_{F}N_{C}(1+x)^{2}}{x}\right\} S_{2}(x)
\end{eqnarray}
\begin{eqnarray}
P^{(1)}_{gq,+-}(x) & = & \left\{ \frac{C_{F}}{36x}\left[ 27C_{F}(4-5x)x-8T_{f}(10+7(x-2)x)\right. \right. \nonumber \\
 &  & \left. \left. \qquad \qquad +2N_{C}(9-3\pi ^{2}(x-1)^{2}-2x(11-x-22x^{2}))\right] \right\} \nonumber \\
 &  & +\left\{ \frac{C_{F}}{6}\left[ 3C_{F}(4+3x)-8N_{C}(6+(x-3)x)\right] \right\} \log x\nonumber \\
 &  & +\left\{ \frac{C_{F}}{3x}\left[ N_{C}(11+8(x-2)x)-4T_{f}(1-x)^{2}\right. \right. \nonumber \\
 &  & \left. \left. \qquad \qquad -3C_{F}(3+2(x-2)x)\right] \right\} \log (1-x)\nonumber \\
 &  & +\left\{ \frac{C_{F}}{2x}\left[ N_{C}+C_{F}(x-2)x\right] \right\} \log ^{2}x\nonumber \\
 &  & +\left\{ -\frac{C_{F}(C_{F}-N_{C})(x-1)^{2}}{x}\right\} \log ^{2}(1-x)\nonumber \\
 &  & +\left\{ -\frac{2C_{F}N_{C}(x-1)^{2}}{x}\right\} \log x\log (1-x)\nonumber \\
 &  & +\left\{ -\frac{C_{F}N_{c}}{x}\right\} S_{2}(x)
\end{eqnarray}
\begin{eqnarray}
P^{(1)}_{gg,++}(x) & = & \left\{ \frac{1}{18x}\left[ 6C_{F}T_{f}(x-1)(x(37+10x)-2)\right. \right. \nonumber \\
 &  & \left. \qquad \qquad +2C_{F}T_{f}(x(21+x(19+23x))-23)\right. \nonumber \\
 &  & \left. \left. \qquad \qquad +N_{C}^{2}(3\pi ^{2}(x(3+x+x^{2})-1)-x(165+103x))\right] \right\} \nonumber \\
 &  & +\left\{ -\frac{2}{3}\left[ 2N_{C}T_{f}(1+x)+6C_{F}T_{f}(2+x)+N_{C}^{2}(x(14+11x)-1)\right] \right\} \log x\nonumber \\
 &  & +\left\{ \frac{4C_{F}T_{f}x(x^{2}-1)+N_{C}^{2}(1+x(6+x(2+(x-8)x)))}{2(1-x)x}\right\} \log ^{2}x\nonumber \\
 &  & +\left\{ \frac{2N_{C}^{2}(1-x(2-2x-x^{3}))}{(x-1)x}\right\} \log x\log (1-x)\nonumber \\
 &  & +\left\{ -\frac{N_{C}^{2}(1+x+3x^{2}+x^{3})}{x}\right\} S_{2}(x)\nonumber \\
 &  & +\left\{ \frac{N_{C}}{9}\left[ N_{C}(67-3\pi ^{2})-20T_{f}\right] \right\} \frac{1}{(1-x)_{+}}\nonumber \\
 &  & +\left\{ N_{C}^{2}\left( \frac{8}{3}+3\zeta (3)\right) -C_{F}T_{f}-\frac{4N_{C}T_{f}}{3}\right\} \delta (1-x)
\end{eqnarray}
\begin{eqnarray}
P^{(1)}_{gg,+-}(x) & = & \left\{ \frac{1}{18x}\left[ 6C_{F}T_{f}(x-1)(x(7+10x)-2)\right. \right. \nonumber \\
 &  & \left. \qquad \qquad +N_{C}^{2}(2(70-3x)x-3\pi ^{2}(1-x+3x^{2}-x^{3}))\right. \nonumber \\
 &  & \left. \left. \qquad \qquad +2N_{C}T_{f}(x(37+x(23x-57))-23)\right] \right\} \nonumber \\
 &  & +\left\{ 2C_{F}T_{f}(1-3x)-N_{C}^{2}\left( 9-13x+\frac{22x^{2}}{3}\right) \right\} \log x\nonumber \\
 &  & +\left\{ \frac{N_{c}^{2}}{2x}\left( 1-x+3x^{2}-x^{3}\right) \right\} \log ^{2}x\nonumber \\
 &  & +\left\{ \frac{2N_{C}^{2}}{x}\left( x^{3}-3x^{2}+x-1\right) \right\} \log x\log (1-x)\nonumber \\
 &  & +\left\{ -\frac{N_{C}^{2}(1+x(2+2x+x^{3}))}{x(1+x)}\right\} S_{2}(x)
\end{eqnarray}

\section{Transversely polarized kernels}

The reference for the transversely polarized kernels is \cite{Vogelsang98}.

\begin{equation}
\Delta _{T}P^{(0)}_{NS^{-}}(x)=\Delta _{T}P^{(0)}_{NS^{+}}(x)=C_{F}\left[ \frac{2}{(1-x)_{+}}-2+\frac{3}{2}\delta (1-x)\right] 
\end{equation}

\begin{eqnarray}
\Delta _{T}P_{NS^{-}}^{(1)} & = & \left\{ \frac{C_{F}}{9}\left[ 20T_{f}-18C_{F}(x-1)+N_{C}(9x-76+3\pi ^{2})\right] \right\} \nonumber \\
 &  & +\left\{ \frac{C_{F}(9C_{F}-11N_{C}+4T_{f})x}{3(x-1)}\right\} \log x\nonumber \\
 &  & +\left\{ \frac{C_{F}N_{C}x}{1-x}\right\} \log ^{2}x\nonumber \\
 &  & +\left\{ \frac{4C_{F}^{2}x}{x-1}\right\} \log x\log (1-x)\nonumber \\
 &  & +\left\{ \frac{2C_{F}(2C_{F}-N_{C})x}{1+x}\right\} S_{2}(x)\nonumber \\
 &  & -\left\{ \frac{C_{F}}{9}\left[ N_{C}(3\pi ^{2}-67)+20T_{f}\right] \right\} \frac{1}{(1-x)_{+}}\nonumber \\
 &  & +\left\{ \frac{C_{F}}{72}\left[ N_{C}(51+44\pi ^{2}-216\zeta (3))-4T_{f}(3+4\pi ^{2})\right. \right. \nonumber \\
 &  & \qquad \qquad \left. \left. +9C_{F}(3-4\pi ^{2}+48\zeta (3))\right] \right\} \delta (1-x)
\end{eqnarray}
\begin{eqnarray}
\Delta _{T}P_{NS^{+}}^{(1)} & = & \left\{ \frac{C_{F}}{9}\left[ N_{C}(3\pi ^{2}-67)+20T_{f}\right] \right\} \nonumber \\
 &  & +\left\{ \frac{C_{F}(9C_{F}-11N_{C}+4T_{f})x}{3(x-1)}\right\} \log x\nonumber \\
 &  & +\left\{ \frac{C_{F}N_{C}x}{1-x}\right\} \log ^{2}x\nonumber \\
 &  & +\left\{ \frac{4C_{F}^{2}x}{x-1}\right\} \log x\log (1-x)\nonumber \\
 &  & +\left\{ \frac{2C_{F}(N_{C}-2C_{F})x}{1+x}\right\} S_{2}(x)\nonumber \\
 &  & -\left\{ \frac{C_{F}}{9}\left[ N_{C}(3\pi ^{2}-67)+20T_{f}\right] \right\} \frac{1}{(1-x)_{+}}\nonumber \\
 &  & +\left\{ \frac{C_{F}}{72}\left[ N_{C}(51+44\pi ^{2}-216\zeta (3))-4T_{f}(3+4\pi ^{2})\right. \right. \nonumber \\
 &  & \qquad \qquad \left. \left. +9C_{F}(3-4\pi ^{2}+48\zeta (3))\right] \right\} \delta (1-x)
\end{eqnarray}

\end{document}